\documentclass{ar-1col-S2O}

\usepackage{graphicx}
\usepackage{amsmath,amssymb}
\usepackage[colorlinks=true,linkcolor=blue,citecolor=blue,urlcolor=blue]{hyperref}

\jname{Annu. Rev. Nucl. Part. Sci.}
\jvol{75}
\jyear{2025}
\doi{10.1146/annurev-nucl-xxxxxx}

\begin{document}

\markboth{Bramante}{Very Heavy Dark Matter}

\title{Very Heavy and Composite Dark Matter: Theory and Experimental Searches}

\author{Joseph Bramante
\affil{Arthur B.~McDonald Canadian Astroparticle Physics Research Institute, 
64~Bader Lane, Queen’s University, Kingston, Ontario K7L~3N6, Canada}
\affil{Department of Physics, Engineering Physics and Astronomy, 
Queen’s University, Kingston, Ontario K7L~3N6, Canada}
\affil{Perimeter Institute for Theoretical Physics, 
31~Caroline Street North, Waterloo, Ontario N2L~2Y5, Canada}
}

\begin{abstract}
Dark matter much heavier than the weak scale remains a comparatively unexplored frontier.  
This review surveys theoretical and experimental developments on very heavy
dark matter, including composite and dissipative formation mechanisms,
multiscatter detection, and astrophysical searches.
\end{abstract}

\begin{keywords}
dark matter, composite dark matter, dissipative sectors, multiscatter detection, neutron stars
\end{keywords}
\maketitle

%Table of Contents
\tableofcontents

\section{Overview, Origins, and Outlook for Heavy Dark Matter}

The nature of dark matter remains a mystery. It is well known that if dark matter is light, then it must couple only very weakly to known matter, lest stars burn faster than they do \cite{Raffelt:1996wa}. But dark matter much heavier than the proton, for which nature provides no natural furnaces, besides the forge of the big bang itself, remains a comparatively unexplored frontier. The aim of this work is to chart out what is known about the origins and detection of very heavy dark matter.

For heavy dark matter that is a single fundamental particle, both cosmological production and detection are well established \cite{Jungman:1995df}. But if dark matter is very heavy, then it is quite plausibly composite, just like atoms, molecules, nuclei, planets, and you and I, dear reader. The idea is not new. It has long been appreciated that the cosmos may contain new bound states of fundamental constituents. The first proposed dark matter composites were variously called quark nuggets \cite{Witten:1984rs}, strange matter \cite{Farhi:1984qu}, and nuclearites \cite{DeRujula:1984axn}. These early studies recognized that if there is confinement or binding forces in a hidden or QCD‐like sector, the universe could be filled with composite dark objects rather than elementary particles. These early composite dark matter studies also began a now well-established tradition: to furnish many different names for functionally identical models of composite dark matter. An incomplete list of dark matter composite names used in recent literature includes q-balls, oscillons, solitons, quasi-solitons, topological defects, axion stars, boson stars, blobs, massive compact halo objects (MACHOs), dark compact objects (DCOs), exotic compact objects (ECOs), fermion stars, macros, asymmetric composites, fermi balls, dark nuclei, and dark quark nuggets. Some of these are distinguishable as being mainly comprised of bosons or fermions; and a proponent of these many names might identify a formation mechanism or Standard Model coupling ($e.g.$ baryon number) as justifying a separate naming convention. Hereafter we will mostly refer to dark matter composites as simply composites, their fundamental constituents as constituents, and when the charges, spins, or masses of the constituents are especially relevant to cosmological formation or present-day detection, we will endeavor to make this known. 

Whatever its form, a steady galactic flow of dark matter moves through our solar system.  With a local density $\rho_X\!\sim\!0.3~\mathrm{GeV/cm^3}$ and speed $v\!\sim\!2.2\times10^7~\mathrm{cm/s}$ \cite{Ou:2023adg}, the mass flux per area is
$
\Phi_X \;\approx\; \rho_X\,v_X\;\sim\;6.6\times10^{10}~\mathrm{GeV\,m^{-2}\,s^{-1}}
\;\Rightarrow\;\sim2\times10^{18}~\mathrm{GeV\,m^{-2}\,yr^{-1}},
$
which is about one reduced Planck mass passing through a square meter per year. This roughly determines the largest dark matter mass (a Planck mass) for which meter-scale human-scale underground dark matter detection experiments are sensitive.

Over the past three decades, direct detection experiments have advanced by many orders of magnitude in sensitivity, led mainly by advances in cryogenic liquid noble and semiconductor detectors.  Yet for this experimental program, some barriers to progress now loom. In the author's opinion, the foremost challenge is an economic and engineering limit: the current multi‐tonne liquid noble detectors such as  PandaX-nT, LZ, XENONnT, and DarkSide‐20k cost on the order of \$50–150 million \cite{Feder:2014dm}, where much of this cost will scale up with the size of the detector, and especially the ultrapure target materials (i.e.~Xenon and Underground Argon \cite{GlobalArgonDarkMatter:2024wtv}). While these experiments have recently scaled up in fiducial volume by a few orders of magnitude per decade \cite{Akerib:2022ort}, such a prodigious size-up cannot continue indefinitely, if only because this growth rate far surpasses the world's overall economic growth \cite{OECD:2018longview}. On top of this, there is an interesting physics background coincidence: within the next few direct detection scale-ups, when costs will begin to give funding agencies pause, these experiments are at the same time going to begin observing a neutrino background. This is the so‐called \emph{neutrino floor} or \emph{neutrino fog} where coherent neutrino–nucleus scattering produces a background indistinguishable from weakly interacting massive particle recoils \cite{Billard:2013qya}. It is important to emphasize that this background, while irreducible given current detector design, is by no means insurmountable: new physics searches can of course operate in the midst of backgrounds \cite{OHare:2016rdm}. So this coincidence, the convergence of escalating experimental costs and the advent of an irreducible neutrino background, only appears to mark a fundamental limit; in practice, it signals that further progress will likely depend on new, inventive, more cost-effective detection techniques.

Given these constraints, several complementary paths forward are being explored. The first is the obvious but costly route of larger instruments - scaling mass and exposure, but at escalating marginal cost given the neutrino fog \cite{Akerib:2022ort}. A second approach trades live mass for \emph{integrated exposure}: paleo‐detectors, aka natural minerals that quietly integrate recoil damage over gigayear timescales and gigaton‐year exposures \cite{Acevedo:2021tbl,Baum:2023cct}. A third strategy uses \emph{astrophysical calorimeters} such as neutron stars and white dwarfs as natural laboratories where dark matter capture and annihilation can manifest as excess heat or dynamical signatures \cite{Bramante:2023djs}. Finally, a nascent fourth path seeks quantum coherence in direct detection: exploiting collective quantum excitations to reach very small effective dark matter couplings \cite{Bhoonah:2019eyo,Monteiro:2020wcb,Arvanitaki:2024taq,Bodas:2025vff}.

The remainder of this review will address some of these detection directions while returning to a broader, older possibility: that dark matter is not a single particle species, but a composite state. We begin with an overview of composite dark matter formation mechanisms.

\section{Composite Dark Matter Formation}
\label{sec:compositeformation}

\begin{figure}[t]
\centering

% --- Top panel: radiative composite formation ---
\includegraphics[width=0.95\linewidth]{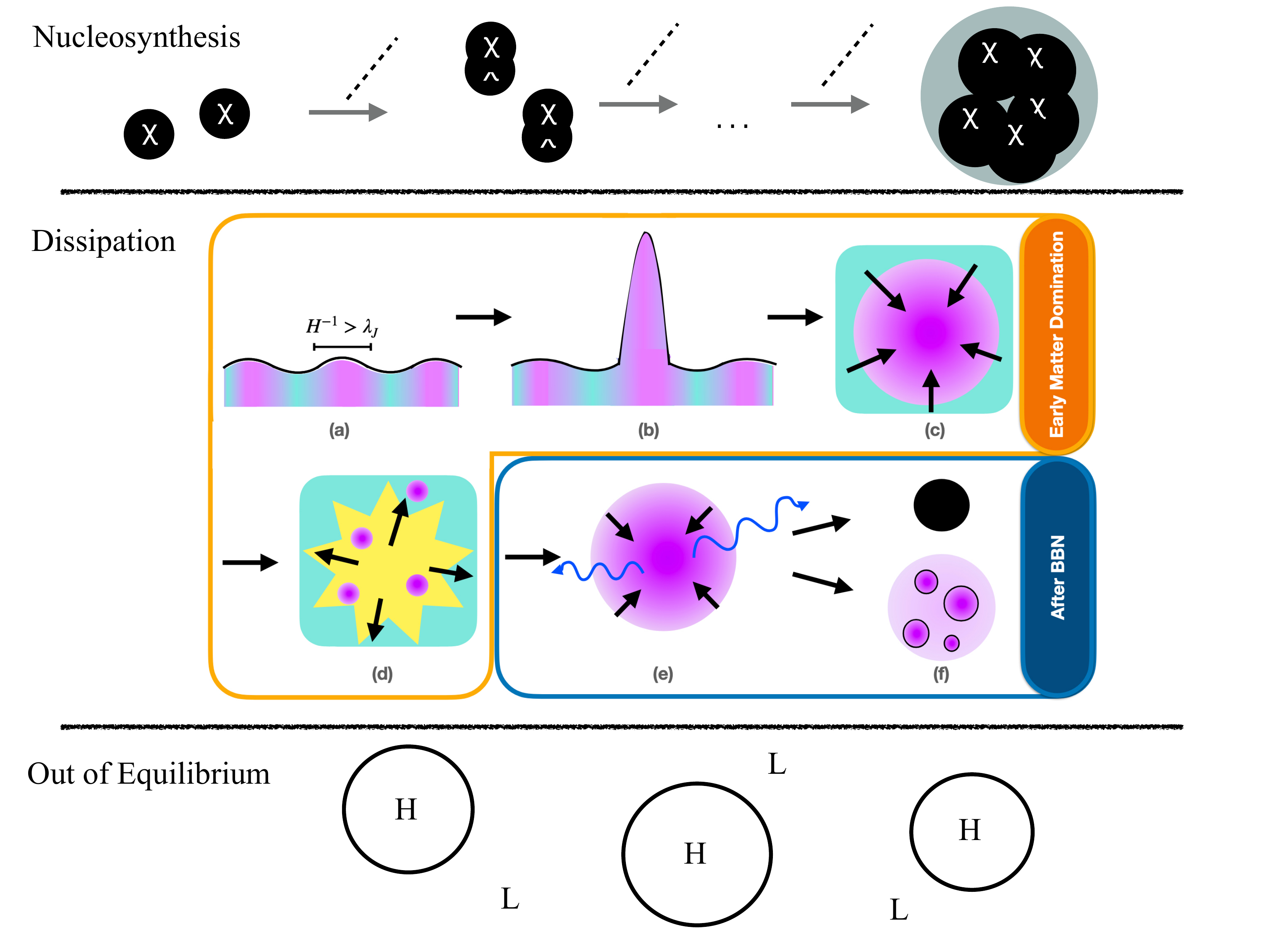}\

\caption{
\textbf{Composite formation pathways for dark matter.}
Top: hierarchical radiative dark nucleosynthesis, where two-body bound states form through mediator emission and merge into larger composites, radiating a scalar $\phi$ at each fusion step.  
Middle: dissipative formation and fragmentation of dark compact objects, adapted from Fig.~1 of~\cite{Bramante:2024pyc}, which shows a sequence of dark halo virialization, radiative cooling, fragmentation, dilution, and low-redshift compact remnant formation.  
Bottom: composite formation in a first-order phase transition, where shrinking bubbles of the high temperature phase trap and compress dark constituents. 
}
\label{fig:composite_formation_scenarios}
\end{figure}

Left to its own devices, dark matter may have done what visible matter does: bind, cool, and complicate the story. We will see that composite dark matter is simple in formulation, but, as with many simple ideas in cosmology, complicated in dynamics. Figure \ref{fig:composite_formation_scenarios} shows three prevalent composite formation pathways, which we now turn to. 

\subsection{Composite Dark Matter Formation from Dark Sector Nucleosynthesis}

Dark matter with an attractive interaction can undergo a process akin to big bang nucleosynthesis, assembling into heavier bound states from lighter constituent fields, if composite synthesis rates exceed the Hubble time after the constituent fields have fallen out of equilibrium with the thermal bath of the universe \cite{Krnjaic:2014xza,Detmold:2014qqa,Detmold:2014kba,Wise:2014ola,Hardy:2014mqa,Bramante:2018qbc,Gresham:2018anj,Ibe:2018juk,Grabowska:2018lnd,Coskuner:2018are,Bai:2018dxf,Bai:2019ogh,Fedderke:2024hfy}. The essential comparison is between the aggregation rate of composite states and the Hubble expansion rate,
\begin{equation}
\Gamma_{\rm syn}(T)\simeq \sum_{i,j} n_i\,\langle\sigma v\rangle_{i+j\to k+\cdots} \gtrsim H(T),
\label{eq:GammaVsH}
\end{equation}
where $\Gamma_{\rm syn}(T)$ represents the total rate for composites of sizes $i$ and $j$ to fuse into a larger composite $k$, $\Gamma_{\rm syn}\simeq\sum_{i,j} n_i\,\langle\sigma v\rangle_{i+j\to k+\cdots}$. Solving this condition for the largest $k$ state formed in a Hubble time, gives a pretty good estimate of the typical size of composites formed \cite{Hardy:2014mqa,Krnjaic:2014xza,Wise:2014jva,Gresham:2017cvl,Gresham:2017zqi}.

A simple model that forms composites includes a heavy asymmetric fermion $X$ interacting via a light scalar mediator $\phi$. Two‐body bound states form by radiative capture, emitting a mediator, and have a ground‐state binding energy
\begin{equation}
{\rm BE}_2 \simeq \frac{\alpha_X^2 m_X}{4},
\label{eq:BE2}
\end{equation}
where $g_X = \sqrt{4 \pi \alpha_X}$ is the scalar-fermion Yukawa coupling, and in the regime that the scalar field has a small mass and temperature $T_\phi, m_\phi \ll {\rm BE}_2 $, capture cross sections scale as $\sigma v \sim \alpha_X^5/\sqrt{m_X^2 T_\phi}$ in a Coulombic regime. On the other hand, this interaction will be suppressed if the mediator is short‐ranged ($m_\phi/(\alpha_X m_X)\gtrsim1$) \cite{Wise:2014jva,Wise:2014ola}. These two-body bound states serve as the seeds for further aggregation.

At large constituent number, fermion-scalar composites enter a saturated regime where the internal number density and mass per constituent stabilize at constant values. The effective constituent mass $\bar m$ inside a saturated composite is set by the equilibrium of Fermi pressure and scalar binding energy and scales as $\bar m_X\simeq m_X - g_X\langle\phi\rangle$, where $\langle\phi\rangle$ is the classical field value inside the bound state \cite{Acevedo:2021kly}. In the saturated regime,
\begin{equation}
\bar{m}_X \simeq 
\left[
  \frac{3\pi\,m_X^{2} m_{\phi}^{2}}{2\,\alpha_X}
\right]^{1/4}.
\label{eq:mbar_sat}
\end{equation}
In this limit the composite radius follows
\begin{equation}
R(N)\simeq\!\left(\frac{9\pi N}{4\,\bar m_X^{3}}\right)^{\!1/3},
\label{eq:saturation}
\end{equation}
and the mean internal density saturates at $n_{\rm sat}\simeq \bar m_X^3/(3\pi^2)$, implying that larger composites grow geometrically, $R\propto N^{1/3}$, until the synthesis rate falls below Hubble.

Loosely bound composites extend this picture to systems whose constituent spacing greatly exceeds the inverse constituent mass, and whose binding energy is much smaller than the constituent rest mass, $E_B \ll m_X \simeq \bar m_X$ \cite{Acevedo:2024lyr}.  They can be understood as dark analogues of nuclei or molecules, with hierarchy $m_X > \Lambda_D > E_B$, where $\Lambda_D^{-1}$ sets the inter‐constituent separation.  For these systems the composite radius scales as
\begin{equation}
R_D \sim N_D^{1/3}\Lambda_D^{-1},
\label{eq:looselybound}
\end{equation}
and synthesis proceeds through $N$–body fusion processes $D_N + D_N \to D_{2N}$ until $\Gamma/H\simeq1$ \cite{Acevedo:2024lyr}.  Their lower densities and larger coherence lengths lead to somewhat different multiscatter signatures, while the reduced binding energy indicates formation at later cosmic times.

In some cases, dark‐sector nucleosynthesis occurs in a \emph{secluded} sector thermally decoupled from the Standard Model.  Composite formation out of equilibrium can then deposit a significant fraction of energy into the binding field, altering its relic abundance and producing a residual $\phi$ population that may contribute to $\Delta N_{\rm eff}$ or constitute a subcomponent of dark matter itself \cite{Bleau:2025klr}.  

Altogether, the composite formation spans a continuum from tightly bound saturated nuggets, where $\bar m_X$ and $n_{\rm sat}$ are constant and growth is geometric, to loosely bound, extended systems where the characteristic scale $\Lambda_D^{-1}$ replaces $\bar m_X^{-1}$ and the assembly timescale is set by delayed, late‐universe nucleosynthesis.  The synthesis condition in Eq.~\eqref{eq:GammaVsH} has been shown to work well to estimate the final composite size for both regimes.

\subsection{Dissipative Formation of Dark Compact Objects}

Dark matter that dissipates defies the usual collisionless picture: once it is cool enough to cool, it can condense into \emph{dark compact objects} (DCOs) and, in some regimes, black holes. We stress here that DCOs are, usually, composite dark matter bound together by the universal attractive force: gravity. Nevertheless, such objects sometimes are called dark compact objects or exotic compact objects, to indicate that they are large enough to be found as nonstandard gravitating bodies in galactic halos (see e.g.~\cite{Liddle:1992fmk,Liebling:2012fv,Giudice:2016zpa,Hippert:2022snq}). Some recent work on dissipative dark matter considered large-scale galactic structures such as dark disks and mirror components \cite{Fan:2013yva,Foot:2013vna} (it is also useful to note that even without dissipation, during an extended period of matter domination, the usual dark sector overdensities are accentuated \cite{Erickcek:2011us}). Later studies have shown that dissipative microphysics can drive collapse on much smaller scales \cite{Buckley:2017ttd,Chang:2018bgx,Ryan:2021dis,Gurian:2022nbx,Bramante:2023ddr,Gemmell:2023trd,Roy:2024bcu,Bramante:2024pyc}.

Dissipative collapse requires that the cooling time fall below the gravitational free-fall time,
\begin{equation}
t_{\mathrm{cool}} \lesssim t_{\mathrm{ff}} = \sqrt{\frac{3\pi}{32G\rho}}\,,
\label{eq:tff}
\end{equation}
so that pressure cannot halt contraction.  Within this basic criterion, several explicit models now show how dark halos can radiate energy and form compact remnants. Simple dark sectors with a long range force can cool efficiently enough for gravitational collapse on sub-galactic scales \cite{Buckley:2017ttd}.  Following a dark electron $+$ dark photon plasma through growth, virialization, and radiative cooling shows conditions under which fragmentation or direct collapse to black holes can occur \cite{Chang:2018bgx,Ryan:2021dis,Bramante:2024pyc}.

A cosmological context for generating DCOs at high abundance arises if the dark sector temporarily dominates the expansion rate, producing an early matter-dominated epoch (EMDE) that amplifies small-scale perturbations \cite{Erickcek:2011us}.  Halos forming at the onset of this phase have characteristic mass
\begin{equation}
M_H \sim \frac{4\pi}{3}\,\rho_{\mathrm{grow}}\,H_{\mathrm{grow}}^{-3},
\label{eq:MH}
\end{equation}
and, if cooling remains efficient, collapse to DCOs or black holes.  A brief later period of energy injection--for instance, from the decay of heavy dark particles--can dilute the background and return the universe to radiation domination before BBN, leaving the collapsed population intact \cite{Bramante:2024pyc}.  In this framework, much or even all of the dark matter may reside in dissipatively formed compact objects. In addition to bremsstrahlung and recombination, an inelastic excited state opens an extra radiative pathway through upscattering and decay \cite{Bramante:2023ddr}. Related work has shown that fermion-scalar bound states can similarly lose pressure support and collapse to black holes \cite{Lu:2024xnb}.

\subsection{Heavy Dark Matter Formed Out of Equilibrium}

The heaviest dark matter candidates may originate not from thermal balance, but from its breakdown during a first order phase transition \cite{Witten:1984rs} or similar out-of-equilibrium phase.  
In confining gauge theories with confinement scale $\Lambda$, the epoch $T \sim \Lambda$ marks a restructuring of the vacuum that alters the population of heavy quarks charged under the gauge theory.  
If the confinement transition is first order, bubble nucleation and latent heat release generate large spatial inhomogeneities.  
The resulting compression of the deconfined phase can trigger a renewed burst of annihilation, in an effect that has been called \emph{thermal squeeze-out}~\cite{Baker:2019ndr,Baker:2021nyl,Asadi:2021pwo,Asadi:2022vkc}.  
Here the annihilation rate will scale with the local (higher) density rather than the global density, accommodating $m_X \gtrsim 10^6~\mathrm{GeV}$ for dark matter produced as a thermally annihilated relic.

When instead the dark sector carries a conserved particle-antiparticle asymmetry (as for the Standard Model, with a quark-antiquark asymmetry of $\sim 10^{-10}$), compression converts the same dynamics into a mechanism of accumulation.  
Following Witten’s original argument for quark nuggets in QCD~\cite{Witten:1984rs}, Bai, Long, and Lu~\cite{Bai:2018dqn} demonstrated that dark quark matter can form stable nuggets whose internal pressure is fixed by the bag constant $B \sim \Lambda^4$, where here $\Lambda$ denotes the confinement scale of the dark gauge group, setting the characteristic energy density of the confining vacuum, while $\mu$ is the chemical potential of the degenerate quark gas inside the nugget.  
The basic assumption of the bag model is that the Fermi pressure from quarks at chemical potential $\mu$ balances the vacuum (bag) pressure $B$, from which we find
\begin{equation}
P_{\mathrm{Fermi}} \simeq \frac{\mu^4}{4\pi^2} = B ,
\end{equation}
which defines the equilibrium energy density inside the nugget and ensures stability against collapse or expansion.  
Their mass and radius approximately follow
\begin{equation}
M_{\rm dQN} \sim B R^3 , \qquad 
R \sim \frac{1}{\Lambda}\left(\frac{M_{\rm dQN}}{M_{\rm Pl}}\right)^{1/3},
\end{equation}
spanning an enormous range from microscopic to astrophysical scales.  
Reference~\cite{Gross:2021qgx} further showed that the same pockets, when dominated by self-gravity, can form degenerate composites (``dark dwarfs") and even primordial black holes. In addition to this, configurations of fields stabilized by attractive forces, which go by the name non-topological solitons or sometimes q-balls \cite{Kusenko:1997si,Amin:2019ums,Hong:2020est,Lu:2024xnb}, are another example of a composite state formed out of equilibrium. Additionally, very-heavy dark matter can be produced dynamically during symmetry-restoration first-order phase transitions \cite{An:2022toi,Freese:2023fcr,Baldes:2018emh}. In such scenarios, the associated stochastic gravitational-wave background—generated by the phase transition or strongly-coupled dark sector dynamics \cite{Schwaller:2015tja} offers a complementary indirect probe of the scale of the hidden sector that produced the dark matter. Finally, it is also possible for dark matter's mass to shift to a higher value, as a consequence of phase transitions in the early universe \cite{Davoudiasl:2019xeb,Bhoonah:2020oov}.

The mechanisms we have reviewed here have a common principle: since confinement or attractive forces can push matter far out of equilibrium, the dark sector can manufacture relics that are both massive and stable, their number densities determined not by Boltzmann suppression of their population after becoming non-relativistic (as with WIMPs), but by the geometry and timing of a first-order phase transition or another out-of-equilibrium process.

\section{Phenomenology of Very Heavy Dark Matter}
\label{sec:phenomenology}

The near-Earth phenomenology of dark matter with masses far exceeding the weak scale is in part shaped by its rarity in terrestrial detectors.  
At $m_X \!\gg\! 10^8\,\mathrm{GeV}$, the flux through terrestrial detectors becomes vanishingly small,
\begin{equation}
\Phi_X = \frac{\rho_X\,v_X}{m_X} 
\simeq 2\,\mathrm{m^{-2}\,yr^{-1}}
\left(\frac{10^{18}\,\mathrm{GeV}}{m_X}\right)
\left(\frac{\rho_X}{0.3\,\mathrm{GeV/cm^3}}\right)
\left(\frac{v_X}{220\,\mathrm{km/s}}\right),
\label{eq:flux}
\end{equation}
so that a meter-scale experiment intercepts of order one particle per year when $m_X \!\sim\! 10^{18}\,\mathrm{GeV}$.  
At such large masses, dark matter can scatter multiple times during its transit through the detector and leave a linear trail of recoils as it passes through the detector. This is the \emph{multiscatter} regime. We note that for masses well below a microgram ($\sim 10^{18}\,\mathrm{GeV}$), terrestrial detectors have largely excluded the parameter space where dark matter has a cross-section large enough to multiscatter, making this correlated-recoil signature a unique hallmark of the highest mass scales.

\subsection{Multiscatter Searches}

The DAMA collaboration was the first to explore multiscatter signatures of heavy dark matter, searching for coincident scintillation bursts across stacked NaI(Tl) crystals in the late 1990s~\cite{Bernabei:1999ui}.  
Their analysis targeted strongly interacting massive particles that might scatter in successive detector layers, producing delayed coincidences between neighboring modules.  
This approach resulted in the first terrestrial scattering limits on dark matter with masses up to $10^{15}\,\mathrm{GeV}$.
The essential principle, that a dark matter particle could leave a sequence of collinear nuclear recoils, has since become one of the basic searches defining heavy dark matter phenomenology.

\subsection{Optical Depth and Multiscatter Kinematics}

For a detector of linear dimension $L_{\rm det}$ and nuclear number density $n_{\rm det}$, the relevant dimensionless quantity for multiscatter is the optical depth \cite{Bramante:2018qbc},
\begin{equation}
\tau \;=\; n_{\rm det}\,\sigma_{X N}\,L_{\rm det},
\label{eq:tau}
\end{equation}
where $n_N$ is the nuclear number density and $\sigma_{NX}$ the nuclear scattering cross-section. The optical depth gives the approximate number of scatters experienced by a dark matter particle during its passage through the detector. Figure \ref{fig:multiscatter} illustrates the relevant scattering regimes.  
When $\tau\!\ll\!1$, the particle typically passes unimpeded, giving the usual single-scatter picture of weakly interacting dark matter.  
When $\tau\!\gtrsim\!1$, each particle interacts about once, and for $\tau\!\gg\!1$ it undergoes multiple collisions, tracing a near-linear track of energy depositions.  
Liquid xenon and argon detectors have $n_{\rm det}\!\simeq\!4\times10^{22}\,\mathrm{cm^{-3}}$, so $\tau\!\sim\!1$ corresponds to $\sigma_{X N}\!\sim2\times\!10^{-23}\,\mathrm{cm^2}$ over a one-meter path.

Because the dark matter is far heavier than the target nuclei, its trajectory is minimally deflected.  
Each scatter produces an angular deflection $\sin\theta_{max} \!\sim\!m_N/m_X$, giving cumulative deflections below $10^{-3}$ radians even for $m_X\!\sim\!10^{10}\,\mathrm{GeV}$.  
The resulting recoil vertices therefore lie nearly collinear, forming a straight-line \emph{recoil track} through the detector, which is a geometric signature that is difficult for background processes to imitate ($e.g.$ neutrons will deflect by large angles during multiscatter transits).

\subsection{Recoil Energetics}

Each elastic collision transfers an energy
\begin{equation}
E_R \simeq \frac{2\mu_{X N}^2v_X^2}{m_N}(1-\cos\theta)
\approx m_Nv_X^2
\sim 100~\mathrm{keV}
\left(\frac{m_N}{100~\mathrm{GeV}}\right)
\left(\frac{v_X}{10^{-3}c}\right)^2,
\label{eq:ER}
\end{equation}
well above nuclear recoil thresholds in modern detectors.  
Over a path length $L_{\rm det}$, the total energy deposition is
\begin{equation}
E_{\rm dep} \sim \tau\,E_R \simeq n_{\rm det}\,\sigma_{X N}\,L_{\rm det}\,E_R,
\label{eq:Edep}
\end{equation}
resulting in total recoil energy depositions well in excess of the $\sim 100 ~{\rm keV}$ expected from single nuclear recoil events.
At large optical depth, the deposited energy approaches quasi-continuous ionization along the path, producing temporally correlated light emission or charge collection extending over microseconds.

\subsection{Elastic Nuclear Scattering}

Analyses of multiscatter dark matter have considered two minimal cross-section parametrizations~\cite{Bramante:2018tos,Acevedo:2024lyr,Cappiello:2020lbk,Digman:2019wdm,Jacobs:2014yca,Davoudiasl:2019xeb}.  

The first assumes \emph{per-nucleon} coupling $\sigma_{X n}$, in which case the per-nucleus scattering rate is enhanced by nuclear coherence,
\begin{equation}
\sigma_{X N} = A^2\left(\frac{\mu_{X N}}{\mu_{X n}}\right)^2\sigma_{X n},
\label{eq:A2}
\end{equation}
where $A$ is the atomic mass number, and the reduced masses of the dark matter with nuclei and nucleons are $\mu_{X N},\mu_{X n}$. This is appropriate for interactions mediated by heavy or contact operators coupling to baryon number or scalar currents, and results in cross sections scaling roughly as $A^4$ for $m_X\!\gg\!m_N$. Such coherence greatly enhances multiscatter rates in high-$A$ targets like xenon. It was noted in \cite{Digman:2019wdm} that if dark matter is a single particle (and not, as also discussed in \cite{Digman:2019wdm}, a composite), there are limits to the total nuclear cross section (essentially the area of the nucleus) one should consider in the per-nucleon scattering formalism. However, later work \cite{Acevedo:2024lyr} discussed in Subsection \ref{subsec:looselybound} has shown that a certain class of loosely bound composites can have effective $A^4$ enhanced scattering, even for very large nuclear cross-sections. 

Another cross-section parameterization is the \emph{geometric} regime, which takes the opposite limit: a \emph{nucleon-independent} or single body cross section $\sigma_X$ that governs the overall interaction probability of the dark matter particle with the detector material, independent of nuclear composition.  
This parametrization applies when scattering proceeds through geometric interactions with composite dark matter, where the fundamental coupling does not coherently add over individual nucleons (aka the hard-sphere scattering limit).  
In this case, $\tau = n_{\rm det}\,\sigma_{NX} L_{\rm det}$ remains the relevant optical depth, but without $A^2$ enhancement, so that
\begin{equation}
    \sigma_{X N} = \sigma_X,
    \label{eq:geo}
\end{equation}
where $\sigma_X$ is the total cross-section of a composite dark matter state.
The distinction between Eqs.~\eqref{eq:A2} and \eqref{eq:geo} thus brackets a plausible range of multiscatter phenomenology, from pointlike per-nucleon couplings to effective macroscopic cross sections.

Finally, we note that there is an important intermediate regime between purely coherent nuclear scattering and geometric whole-composite scattering. When the momentum transfer is sufficiently large to resolve the internal structure of the dark matter state, the interaction transitions to incoherent scattering on individual constituents. This intermediate kinematic regime is particularly relevant for loosely bound composites, and we detail its phenomenology and cross-section scalings in Section \ref{subsec:looselybound}.

\subsection{Experimental Morphology and Astrometric Prospects}

In the multiscatter regime ($\tau\!\gtrsim\!1$), each transit produces a chain of correlated recoils that can be time-ordered by scintillation or charge collection.  
For $\tau\!\sim\!1$--$10$, discrete recoils are likely separable; for $\tau\!\gg\!10$, they merge into a single extended pulse with a characteristic asymmetric time profile, as modeled in large-volume detectors~\cite{Bramante:2018tos}.  
Because these events occupy straight trajectories through the fiducial volume, reconstruction of their direction and duration allows an estimate of the incident velocity vector $v_X$ and directionality of the galactic dark matter wind \cite{Bramante:2019yss}.

\begin{figure}[t]
\begin{minipage}[t]{0.48\textwidth}
\centering
\includegraphics[width=\linewidth]{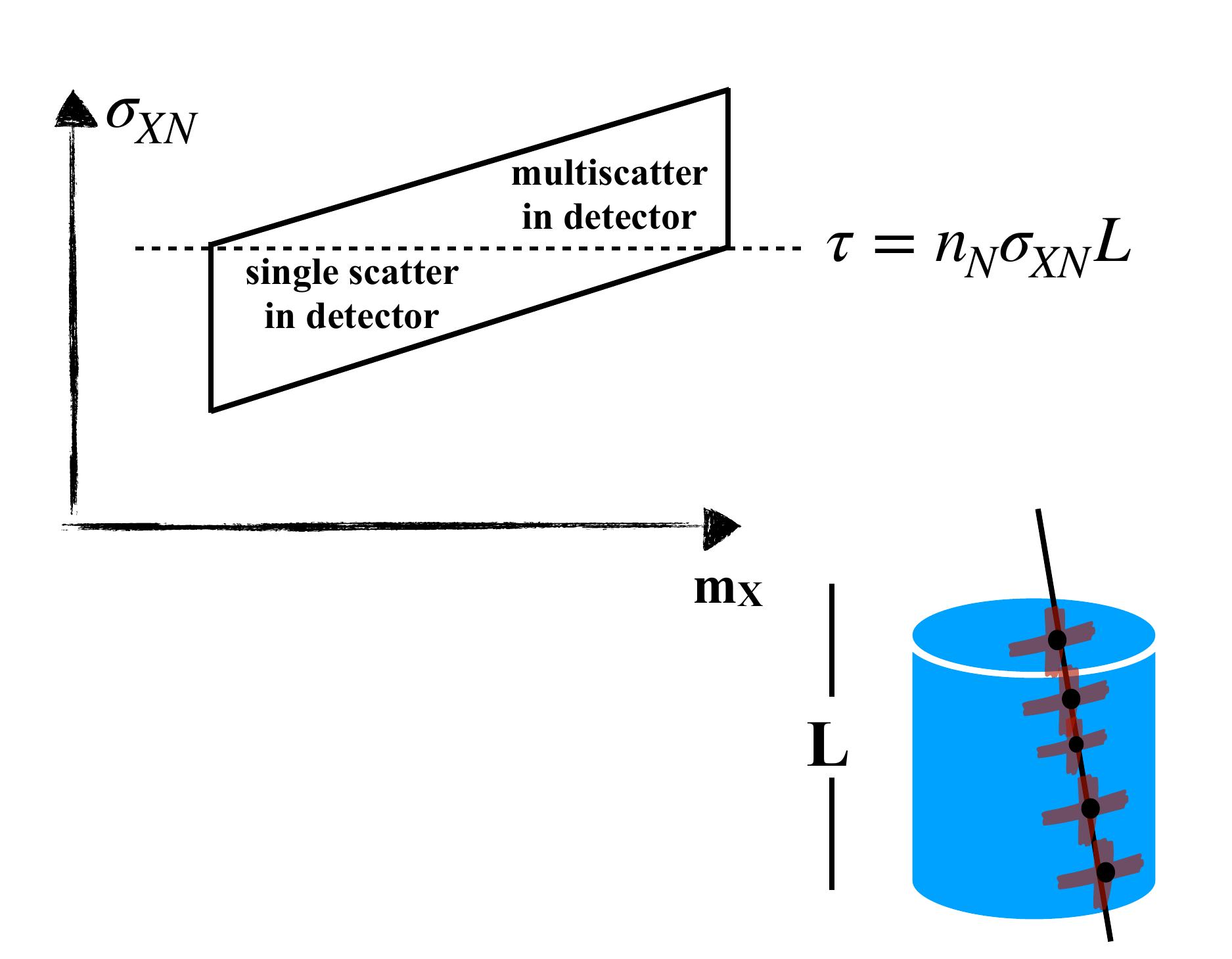}
\end{minipage}
\hfill
\begin{minipage}[t]{0.48\textwidth}
\centering
\includegraphics[width=\linewidth]{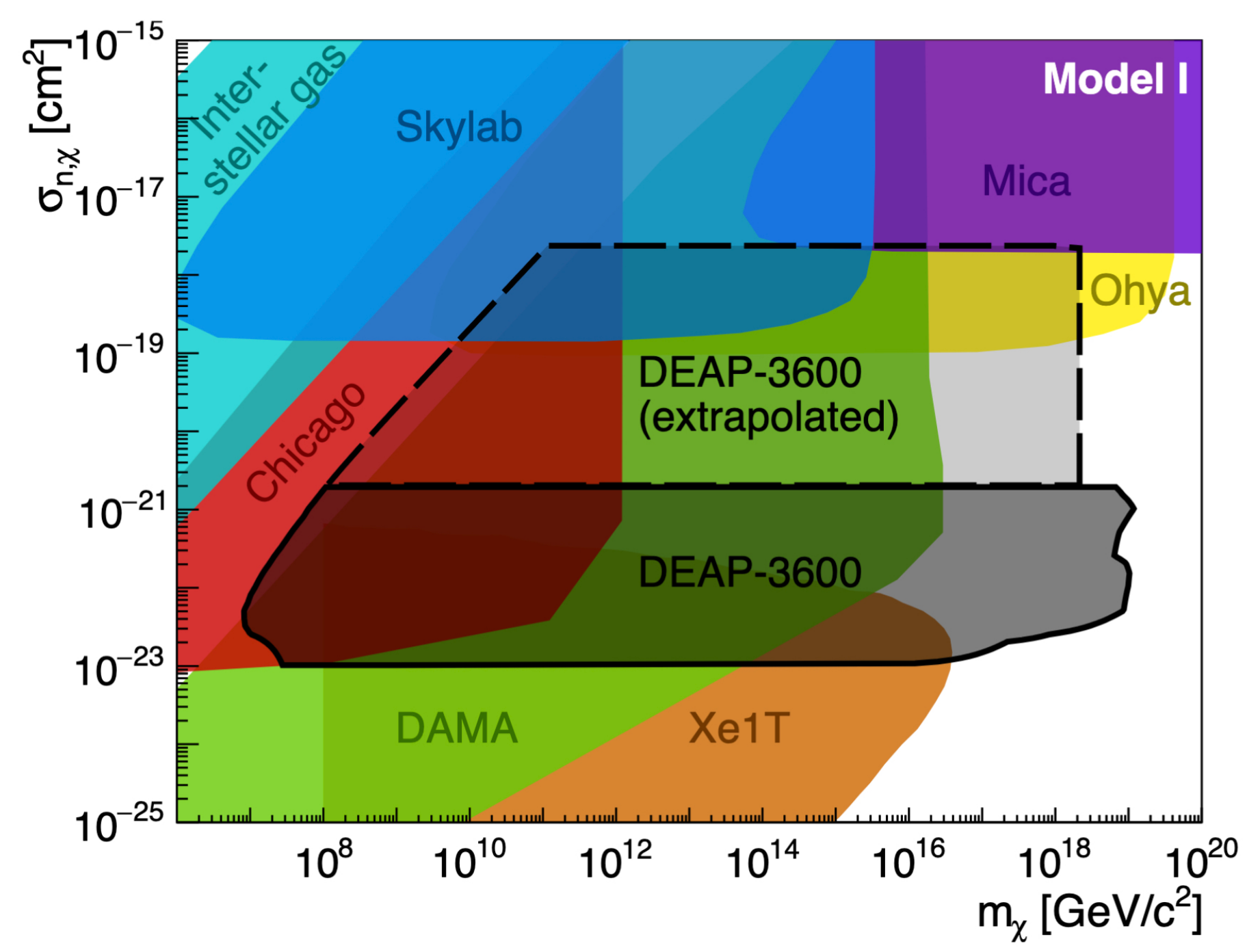}
\end{minipage}
\caption{
\textbf{Multiscatter phenomenology of very heavy dark matter.}
Left: schematic of multiple nuclear recoils aligned along a single linear trajectory for $m_X\!\gg\!m_N$, producing a correlated light or charge track~\cite{Bramante:2018tos}. 
Right: exclusion region from the DEAP-3600 multiscatter search~\cite{DEAPCollaboration:2021raj}, constraining dark matter masses up to $10^{19}\,\mathrm{GeV}$. The enclosed region is bounded below by the minimum cross-section required to produce a detectable multiscatter track using the pulse-timing discrimination analysis at DEAP-3600, above by energy attenuation in the Earth's overburden, and on the right by the local dark matter flux. Variations in the terminus of the right boundary depend on DEAP-3600 sensitivity to different energy thresholds/high-velocity phase space of dark matter.}
\label{fig:multiscatter}
\end{figure}

Figure~\ref{fig:multiscatter} illustrates this phenomenology: the left panel schematically depicts collinear nuclear recoils along a track, which will be collinear for elastic recoils in the $m_X\!\gg\!m_N$ limit, while the right panel shows the DEAP--3600 search region and some current constraints~\cite{DEAPCollaboration:2021raj}.  
A comprehensive survey of existing and projected multiscatter searches across detector classes follows in the next section.

Finally, we note that because the heavy dark matter trajectory is nearly straight, the sequence of time-tagged scatters defines a track whose direction reconstructs the local dark matter wind in the Galactic frame.  
Track reconstruction in large liquid scintillator detectors can recover the velocity of individual transits to better than $\Delta v/v\sim10^{-3}$~\cite{Bramante:2019yss}.  
With $\mathcal{O}(10)$ detected events, such detectors could independently begin mapping the dark matter velocity dispersion, escape speed, and anisotropy of the local halo.  

\subsection{Scattering of Nuclei Inside a Composite}
\label{subsec:insidecomposite}

A minimal way for Standard Model nuclei to interact with a large composite is through a Yukawa portal between the composite’s binding field $\phi$ and nucleons \cite{Acevedo:2020avd,Acevedo:2021kly},
\begin{equation}
\mathcal{L}_D = \bar X(i\!\not\!\partial-m_X)X + g_X\,\bar X \phi X + \frac{1}{2}(\partial\phi)^2 + \frac{1}{2}m_\phi^2 \phi^2,
\qquad
\mathcal{L}_{n} = g_n\,\bar n \phi n,
\label{eq:LD-portal}
\end{equation}
where in this Lagrangian $X$ is the dark matter fermion field, $m_X$ is the dark matter fermion mass, $g_X$ is the fermion Yukawa coupling to $\phi$, and $g_n$ is the Yukawa coupling of $\phi$ to SM nucleons $n$.
In saturated composites (large constituent number), the binding field attains a classical interior value $\langle\phi\rangle \simeq m_X/g_X$, so a nucleus of mass $m_N$ and atomic number $A$ experiences an effective scalar potential $\langle\varphi\rangle \equiv A g_n\langle\phi\rangle$ across the composite boundary. For nonrelativistic boundary crossing the nucleus gains (or loses) kinetic energy
\begin{equation}
\Delta E \simeq A\,g_n\,\frac{m_X}{g_X},
\label{eq:deltaE}
\end{equation}
and, for a short-enough range (the skin thickness of the composite is $m_\phi^{-1}$), the acceleration can occur over a time short enough to produce Migdal ionization in noble targets \cite{Acevedo:2021kly}. 

\begin{figure}[t]
    \centering
    \includegraphics[width=0.99\textwidth]{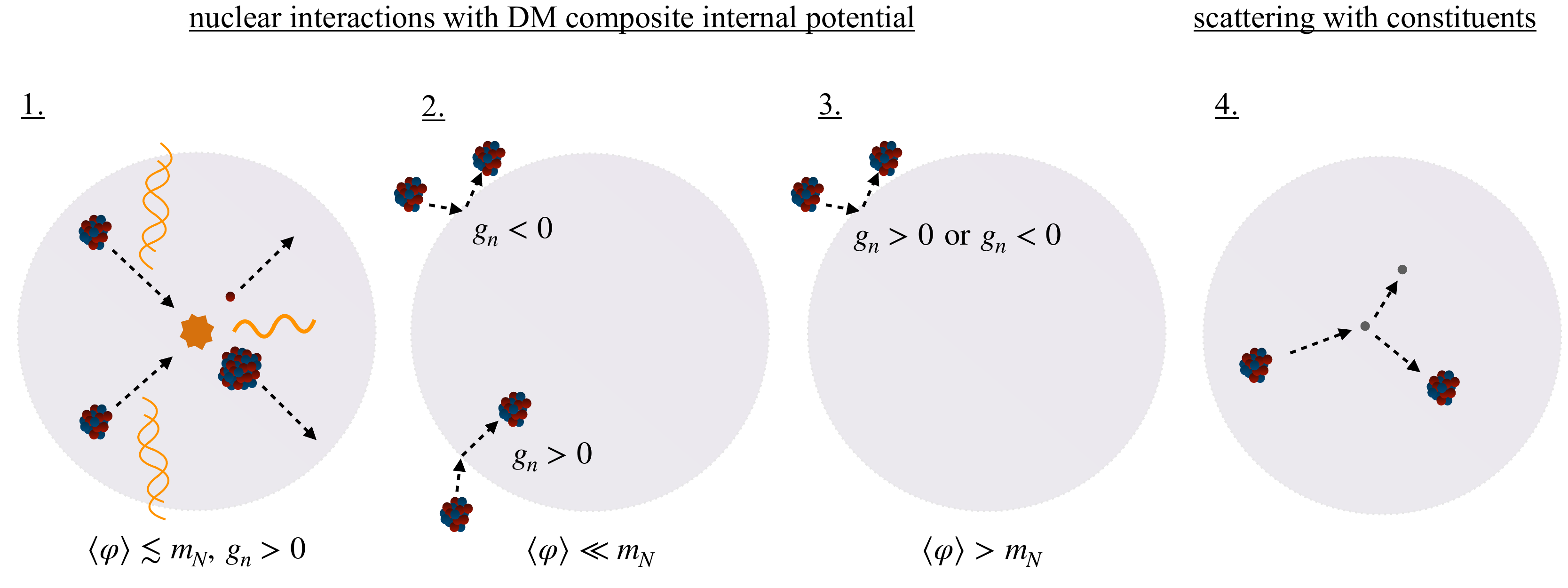}
    \caption{
    \textbf{Schematic regimes of composite dark matter scattering in detectors.}
    Four conceptual panels are shown from left to right:
    (1.) \textit{Whole-composite boundary acceleration:} an incoming dark composite interacts coherently with a nucleus, transferring momentum to the entire bound state and producing nuclear recoil radiation;
    (2.) \textit{Soft boundary recoils and Migdal emission:} partial or glancing interactions excite electrons and emit soft bremsstrahlung photons, leading to correlated electronic signals along the track;
    (3.) \textit{Reflection or saturation regime:} for large effective mediator field or large $|\langle\varphi\rangle|$, the scattering becomes surface-dominated and the momentum transfer saturates near the geometric limit;
    (4.) \textit{Loosely bound composite regime:} the probe resolves individual constituents within an extended dark bound state, producing incoherent constituent-level scattering as described in Sec.~\ref{subsec:looselybound}.
    Adapted from Fig.~1 of \cite{Acevedo:2021kly}, with the final panel illustrating the loosely bound composite regime explored in \cite{Acevedo:2024lyr}.}
    \label{fig:composite_scattering_schematic}
\end{figure}

These dynamics of nuclei interacting at the boundary of a large composite can be organized into four regimes: (i) modest attractive potentials $\langle\varphi\rangle\!\lesssim\!m_N$ yield nuclear acceleration with bremsstrahlung, as well as possible nuclear fusion if the composite traverses dense target environments like white dwarfs (detailed further in Section 4.2); (ii) small $|\langle\varphi\rangle|\!\ll\!m_N$ produce soft boundary recoils and Migdal electrons in underground detectors; (iii) very large $|\langle\varphi\rangle|\!\gtrsim\!m_N$ effectively reflect nuclei, achieving the geometric limit given in Eq.~\eqref{eq:geo}; and (iv) direct scattering on constituents inside the composite. This final regime is usually Pauli–suppressed in saturated composites, but can be the predominant interaction for loosely-bound composites, reviewed in the next subsection. All four of these composite interaction regimes are illustrated in Figure \ref{fig:composite_scattering_schematic}. Finally, we note that for whole-composite scattering (no interior access), microscopic solid-state searches have proposed imaging damage tracks in geological targets \cite{Acevedo:2021tbl,Baum:2023cct,Ebadi:2021cte,Drukier:2018pdy,Fung:2025cub}.

\subsection{Loosely Bound Composites and Scattering on Constituents}
\label{subsec:looselybound}

Loosely bound composites occupy a different structural limit: their constituent mass exceeds the inter-constituent scale, and both far exceed the binding energy per constituent, e.g.
\begin{equation}
m_d \;>\; \Lambda_D \;>\; E_B, 
\qquad 
R_D \sim N_D^{1/3}\,\Lambda_D^{-1},
\label{eq:LB-hierarchy}
\end{equation}
with $R_D$ the composite radius and $\Lambda_D^{-1}$ the inter-constituent spacing ($i.e.$ nuclear or molecular binding length, depending on the model), where the constituent mass $m_d$ is defined as the mass of constituents when they are bound inside the dark matter composite (i.e. $m_d$ will be less than the bare mass of the constituents, $m_d + E_B$). 

In this regime, detector kinematics interpolate across four scattering modes as momentum transfer increases \cite{Acevedo:2024lyr}: (i) pointlike coherent scattering (small $R_D q$), (ii) coherent but form-factor–suppressed scattering off the extended density, (iii) incoherent scattering on single constituents (with possible internal excitations), and (iv) spallation when $E_B$ is sufficiently small. A useful empirical upshot is that coherent nucleus–composite scattering can scale very steeply with target mass number, up to $\sim A^4$ in representative models, dramatically enhancing multiscatter rates in high-$A$ media.

For estimates, the geometric growth of $R_D$ in Eq.~\eqref{eq:LB-hierarchy} increases the composite form factor’s role at moderate $q$, shifting events from pointlike to extended-coherent response as $m_X v$ rises. At still larger $q$ (set by $\mu_{A d}$), nuclei resolve and scatter on individual constituents; rates then depend on a constituent–nucleon cross section $\sigma_{nd}$ and the composite’s internal state (including excitation channels). The taxonomy and detector-facing scalings for these regimes, including the conditions under which constituent scattering dominates, are laid out in \cite{Acevedo:2024lyr}. 

This constituent scattering picture illustrated in Figure \ref{fig:composite_scattering_schematic} also clarifies how “loosely bound” composites differ from saturated nuggets tied to Eq.~\eqref{eq:LD-portal}: coherence lengths and reduced densities result in many very soft recoils against Standard Model nuclei or electrons per transit through a detector (potentially $\ll$\,keV each but with large integrated $E_{\rm dep}$). This signature space is orthogonal both to single-site WIMP recoil searches and also are somewhat different from the recently established multiscatter analyses. While this topic is still under exploration, Ref.~\cite{Acevedo:2024lyr} provides worked examples and projected energy depositions in argon and xenon. 

\section{Terrestrial Searches for Very Heavy Dark Matter}

\subsection{Multiscatter Dark Matter Searches at Underground Direct Detection Experiments}
\label{subsec:multiscatter}

Building on the phenomenology outlined in Sec.~\ref{sec:phenomenology}, terrestrial experiments have now performed dedicated searches for dark matter that would scatter multiple times while traversing the detector.  In this \emph{multiscatter} regime, the optical depth $\tau = n_{\rm det}\sigma_{X N}L_{\rm det}$ exceeds unity, and the resulting sequence of near-collinear nuclear recoils leaves an extended scintillation or ionization trail.

\begin{figure}[t]
\centering
% --- Top row: square panels ---
\begin{minipage}[t]{0.48\textwidth}
\centering
\includegraphics[width=\linewidth]{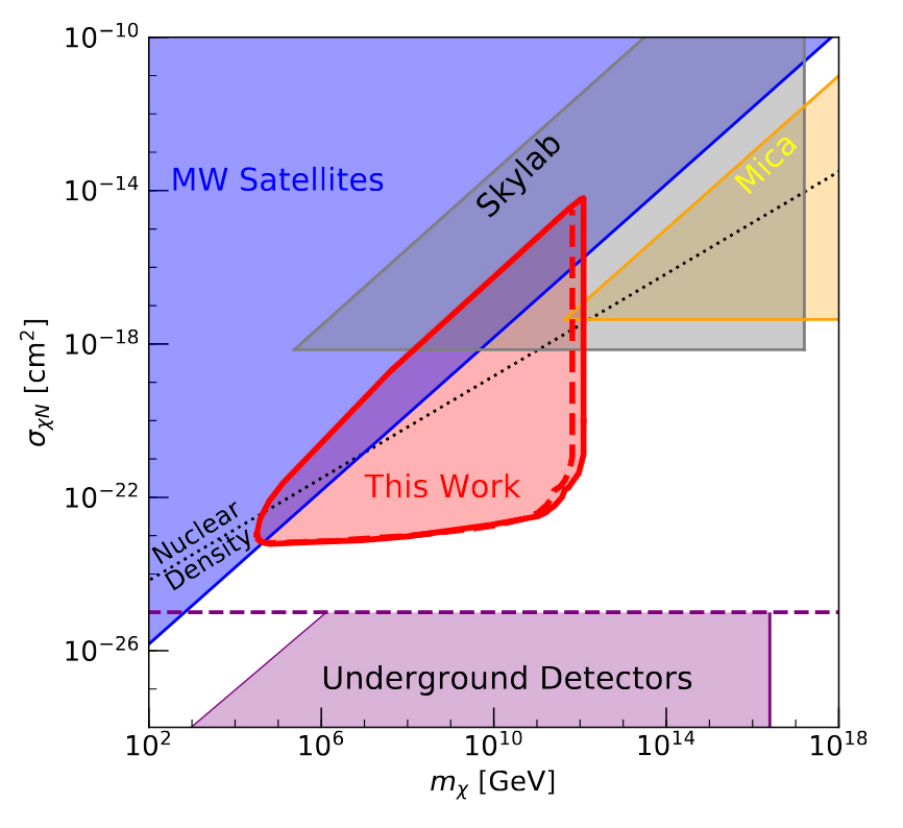}
\end{minipage}
\hfill
\begin{minipage}[t]{0.48\textwidth}
\centering
\includegraphics[width=\linewidth]{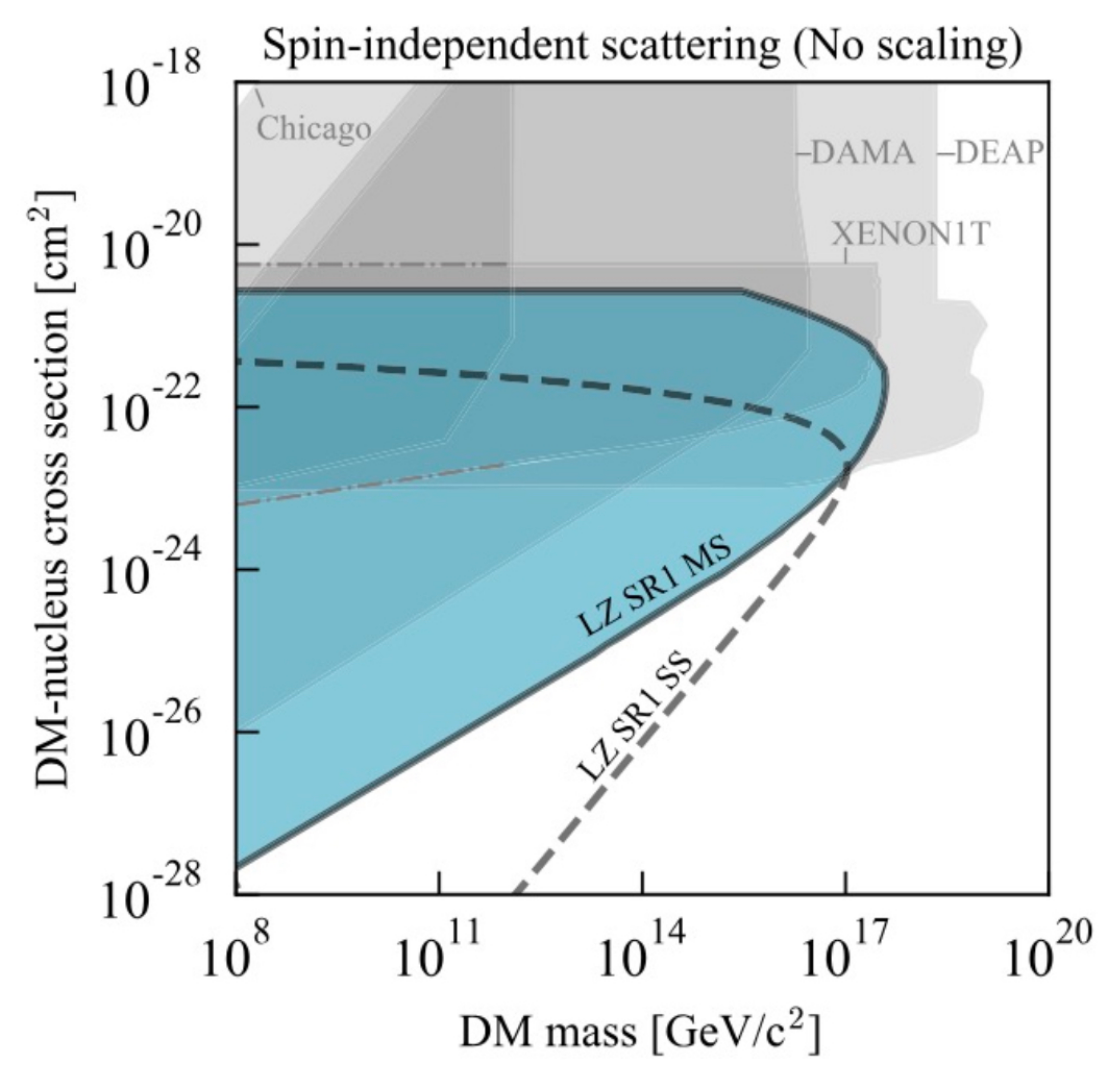}
\end{minipage}

%\vspace{6pt}

% --- Bottom row: landscape panels ---
%\begin{minipage}[t]{0.48\textwidth}
%\centering
%\includegraphics[width=\linewidth]{figures/cdms_recast_panel.pdf}
%\end{minipage}
%\hfill
%\begin{minipage}[t]{0.48\textwidth}
%\centering
%\includegraphics[width=\linewidth]{figures/xenon1t_panel.pdf}
%\end{minipage}

\caption{
\textbf{Some multiscatter and high-mass dark matter limits from terrestrial detectors.}
Left: Chicago EJ-301 coincidence search~\cite{Cappiello:2020lbk};
Right: LZ multiscatter search~\cite{LZ:2024psa}. The relatively flat lower bound of the EJ-301 search is determined by a threshold for total nuclear recoil deposition in a liquid scintillator detector, while the LZ SR1 multiscatter search has a more characteristic $1/ m_X$ scaling of a single-recoil sensitive detector. 
}
\label{fig:terrsearches}
\end{figure}

As already mentioned, the first explicit search for such events was conducted by the DAMA collaboration in 1999~\cite{Bernabei:1999ui}, which sought coincident scintillation bursts across adjacent NaI(Tl) crystals from slow, heavy dark matter.  Three recently published large-volume detection studies—XENON1T~\cite{XENON:2023iku}, LZ~\cite{LZ:2024psa}, and DEAP-3600~\cite{DEAPCollaboration:2021raj}—have each undertaken high-mass analyses specifically targeting multiple scatters within single particle tracks.  
Because these detectors combine tonne-scale target masses with nanosecond timing, they can identify extended pulses consistent with sequential nuclear recoils.  
No events have been observed, leading to the strongest constraints to date on both per-nucleon and nucleon-independent scattering models:
\begin{itemize}
  \item The \textbf{XENON1T} multiscatter window excludes $\sigma_{X N}\!\gtrsim\!10^{-23}\,\mathrm{cm^2}$ for $m_X \!\sim\!10^{12}$–$10^{17}\,\mathrm{GeV}$. (Fig.~\ref{fig:terrsearches})
  \item The \textbf{LZ} study explicitly modeled a single scatter and multiscatter window, again obtaining sensitivity up to $m_X \!\sim 10^{17}\,\mathrm{GeV}$. (Fig.~\ref{fig:terrsearches})
  \item Multiscatter search analyses at the \textbf{PandaX-4T} experiment \cite{PandaX:2024qfu} are ongoing.
  \item \textbf{DEAP--3600} obtained a wide sensitivity to high cross-sections in argon, with complementary nuclear response and timing discrimination, and has the highest mass reach of any terrestrial multiscatter dark matter search, up to the Planck mass. (Fig.~\ref{fig:multiscatter})
\end{itemize}
Together these analyses have sought a large portion of parameter space for composite heavy dark matter below the Planck mass.

A simple strategy has been developed in~\cite{Cappiello:2020lbk} for lower mass multiscatter composites. Ref~\cite{Cappiello:2020lbk} operated two EJ-301 scintillator modules separated by fifty cm to search for dark matter crossing both detectors with a microsecond-scale time-of-flight. This \emph{Chicago} coincidence experiment ruled out nucleon-independent cross sections $\sigma_X\!\gtrsim\!10^{-22}\,\mathrm{cm^2}$ for $m_X\!\lesssim\!10^{12}\,\mathrm{GeV}$, illustrating that scintillating coincidence detectors can efficiently test the geometric scattering regime. The heavy dark matter analysis of \cite{Bhoonah:2020dzs} combined laboratory and astrophysical methods to obtain complementary reach for strongly interacting and composite dark matter. In that work, prior data from CDMS-I, CRESST, and XQC were reinterpreted to obtain geometric cross-section limits that account for energy loss through the atmosphere and surrounding overburden. At large cross sections, attenuation suppresses the flux of dark matter reaching underground detectors, while shallow-site and high-altitude experiments remain sensitive to interactions near the geometric limit. Ref.~\cite{Bhoonah:2020dzs} also showed that interstellar gas cloud heating probes a complementary composite parameter space.

Underground detectors, coincidence arrays, and recasts of existing data now together constrain 
dark matter across nearly twenty orders of magnitude in mass and six in cross section.  
Future efforts combining low-threshold sensors with larger active volumes will be needed to extend sensitivity at the sub-Planck mass frontier of composite dark matter.

\subsection{Heavy Dark Matter Searches at Neutrino Detectors and Other Large-Volume Experiments}

To date, there have been no dedicated multiscatter searches for very heavy dark matter by Super-Kamiokande, SNO+, IceCube, or other large volume neutrino detectors, though efforts towards achieving this are ongoing. The closest published analogue is IceCube’s nonrelativistic magnetic monopole analysis, which implements slow-track triggers and background rejection for sub-Cherenkov and mildly luminous trajectories, setting stringent flux limits on monopoles and demonstrating the feasibility of rare, slow object reconstruction in a km$^3$ Cherenkov array \cite{IceCube:2014xnp}. In parallel, it has been shown that large liquid-scintillator detectors (Borexino/SNO+/JUNO scale) could identify dark matter through \emph{multiscatter} linear tracks if the per-nucleon cross section and optical depth are sufficient; this requires only modest trigger logic targeting temporally extended, collinear light emission \cite{Bramante:2018tos}.

A distinct and potentially more powerful avenue arises for \emph{composite} dark matter whose constituents source a binding field that accelerates traversing Standard-Model nuclei \emph{inside} the composite. For this kind of composite interaction \cite{Acevedo:2020avd}, reviewed above in Subsection \ref{subsec:insidecomposite}, an incident nucleus experiences a coherent potential while it crosses the composite, acquiring kinetic energy that scales linearly with the binding-field's potential and the binding field's coupling to nucleons; subsequent in-medium processes (e.g., bremsstrahlung, plasma emission, or even fusion when kinematically allowed) yield observable secondary photons in the keV–GeV range along the transit. Crucially, the yield scales with the scalar-nucleon coupling (denoted $g_n$ in this review) and with the composite's overall size and structure rather than exhibiting an $A^2$ enhancement. As detailed in Section \ref{subsec:insidecomposite} and illustrated in the first panel of Figure \ref{fig:composite_scattering_schematic}, this is because the interaction is not a hard scatter resolving individual nucleons, but rather a coherent acceleration of the nucleus through the entire composite's internal potential. The resulting signature in large instruments is a bright, non-relativistic track composed of time-correlated photon and possibly hadronic sub-cascades, distinct from through-going muons by speed, photon spectrum, and (for very heavy dark matter) monotonicity of secondary production along the length of the track. 

For liquid scintillator, secondary photons result in a prompt, track-like signature with microsecond-scale timing across tens of centimeters; for water/ice Cherenkov arrays, the same processes can lift sub-Cherenkov-speed dark matter transits into a detectable regime by generating above-threshold quantities of photons. The IceCube slow-particle trigger demonstrated in \cite{IceCube:2014xnp} is directly applicable to such composite-induced radiation, and sensitivity projections indicate that IceCube- and SNO+/JUNO-scale instruments can probe regions of $(m_X, g_n)$ that are inaccessible to standard underground single-recoil analyses when composite-internal nuclear is the predominant dark matter interaction process \cite{Acevedo:2020avd,Bramante:2018tos}. 

In summary, large-volume detectors already possess (i) slow-track triggering and reconstruction validated on monopole searches \cite{IceCube:2014xnp} and (ii) the timing and segmentation needed to follow extended, correlated light profiles \cite{Bramante:2018tos}. What remains is a more dedicated set of large-volume analyses targeting composite-induced radiation along heavy dark matter tracks \cite{Acevedo:2020avd,Acevedo:2021kly}. Developing such searches is a clear opportunity for near term progress in the very heavy and composite dark matter regime.

\subsection{Heavy Dark Matter Searches Using Plastic Etch Detectors and Mica}
\label{subsec:mica-plastic}

Solid-state damage track detectors provide a complementary path to very heavy or composite dark matter. Their virtue is simple: a slow, massive particle depositing large localized energy along a straight path can leave a permanent lattice defect, later revealed by chemical etching (plastics) or by cleaving, etching, and other forms of microscopic inspection (mica). Historically, these methods were developed for high-$Z$ cosmic rays and magnetic monopoles; in the modern dark matter context, macro constraints and expected fluxes were reexamined in \cite{Jacobs:2014yca}, motivating new dedicated analyses that account for overburden attenuation, detector thresholds, and long-term annealing.

For a projectile of speed $v\!\sim\!10^{-3}c$ moving through material of density $\rho$, the restricted energy loss is approximately
\begin{equation}
\left(\frac{dE}{dx}\right) \sim \rho\,\sigma_{\rm eff}\,v^2,
\label{eq:dedx2}
\end{equation}
where $\sigma_{\rm eff}$ represents the relevant cross section. Track formation requires $(dE/dx)$ above the latent-damage threshold of the detector medium. Two limiting parameterizations detailed above in Eqs.~\eqref{eq:A2},\eqref{eq:geo} give customary regimes of interest: a per-nucleon model with $\sigma_{\rm eff}=A^2(\mu_{X N}/\mu_{X n})^2\sigma_{X n}$, and a nucleon-independent (geometric) model with a single effective area $\sigma_{\rm eff}=\sigma_X$, independent of $A$. The corresponding flux attenuation in the atmosphere or rock overburden over some overburden length $z$ is governed by (integrating Eqn.~\eqref{eq:dedx2} with $E = m_X v^2/2$)
\begin{equation}
    E_{\rm f}
\simeq E_{\rm i}\,
\exp\!\left[
-\frac{2}{m_X}
\int_{\rm path} \! dz \;
\sum_{A} n_N(z)\,\frac{\mu_{X N}^2}{m_N}\,\sigma_{\rm eff}
\right],
\label{eq:EnergyAttenMaster}
\end{equation}
where the sum is over nuclei/atoms with number density $n_N$ in the overburden.

Acceptance in solid-state detectors scales with detector area and exposure time (for plastics) or with area and geologic age (for mica), providing different but complementary integration timescales. Etched plastics such as CR-39 and Lexan record tracks when the local energy loss exceeds the chemical etch threshold, producing etch-pit cones where the track etch rate $v_T$ exceeds the bulk rate $v_B$. Building on these techniques, \cite{Bhoonah:2020fys} used published exposures from cosmic ray and monopole search programs \cite{Skylab,Orito:1990ny,Starkman:1990nj} to constrain heavy and composite dark matter in both parameterizations above, including overburden and detector-threshold effects. Etched plastics probe large $\sigma_{\rm eff}$ values inaccessible to underground experiments once attenuation dominates, providing the leading limits in the geometric (nucleon-independent) regime and complementary reach in the coherent per-nucleon case. Their scalability and surface deployment potential make them an efficient probe of heavy dark matter flux at high cross section.

Ancient mica dark matter searches \cite{Price:1986ky,SnowdenIfft:1995ke} offer a different compelling approach. Natural muscovite mica may retain damage tracks for gigayear timescales, acting as a billion-year exposure detector for rare, heavy particles. Early work in \cite{Price:1986ky,SnowdenIfft:1995ke} introduced systematic selection of samples by geologic age and temperature history, controlled cleaving and etching, coincidence matching of tracks on opposite surfaces, and modeling of fission fragment and $\alpha$-recoil backgrounds. In the modern heavy dark matter context, Ref.~\cite{Acevedo:2021tbl} showed that the null observation of straight, through-going tracks in the past studies translates into leading bounds on both per-nucleon and nucleon-independent cross sections across masses from $10^15$ to $10^{25}\,\mathrm{GeV}$—precisely the region where dark matter fluxes are tiny but individual crossings could deposit GeV$/$cm energy along their path through earth. Ancient mica’s key advantages are its large effective exposure (area $\times$ age) and its ability to record distinctive straight, low-deflection tracks characteristic of $m_X\!\gg\!m_N$ kinematics, in parallel sheets of layered muscovite. Although the thermal track annealing models (which are important, since annealed tracks are erased) were calibrated in \cite{Price:1986ky,SnowdenIfft:1995ke}, it would be good in the modern context to revisit annealing in mica to reliably reconstruct track survival probability over geologic times, for future searches.

\begin{figure}[t]
 \centering
% --- Top row: Plastic panels ---
%\begin{minipage}[t]{0.48\textwidth}
%\centering
%\includegraphics[width=\linewidth]{figures/plastic_perN.pdf}
%\end{minipage}
%\hfill
%\begin{minipage}[t]{0.48\textwidth}
%\centering
%\includegraphics[width=\linewidth]{figures/plastic_geom.pdf}
%\end{minipage}

%\vspace{6pt}

% --- Bottom row: Mica panels ---
\begin{minipage}[t]{0.48\textwidth}
\centering
\includegraphics[width=\linewidth]{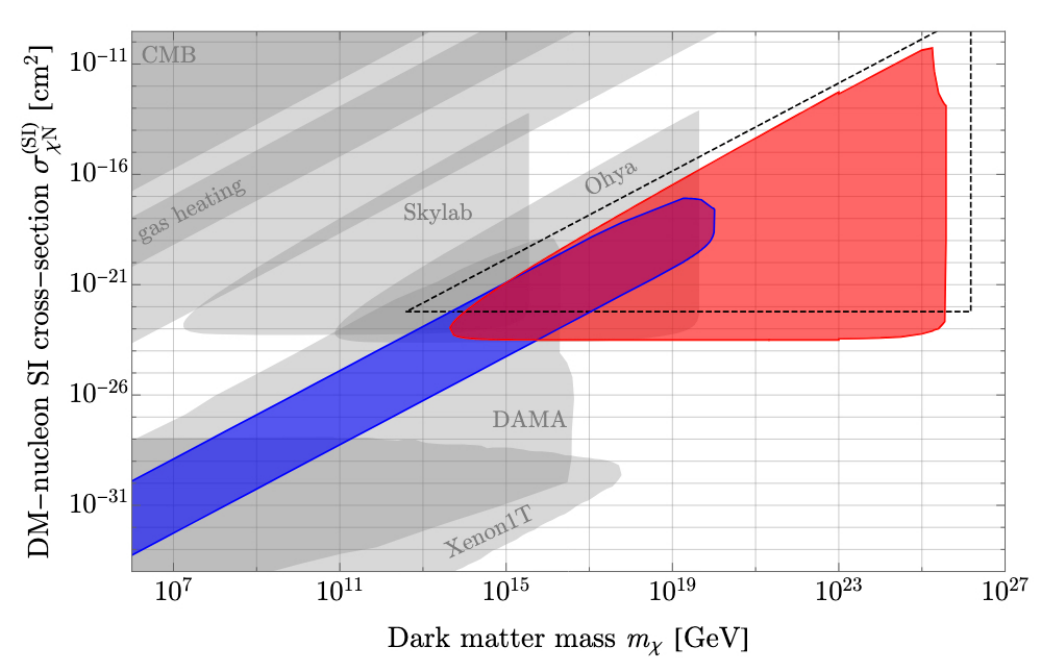}
\end{minipage}
\hfill
\begin{minipage}[t]{0.48\textwidth}
\centering
\includegraphics[width=\linewidth]{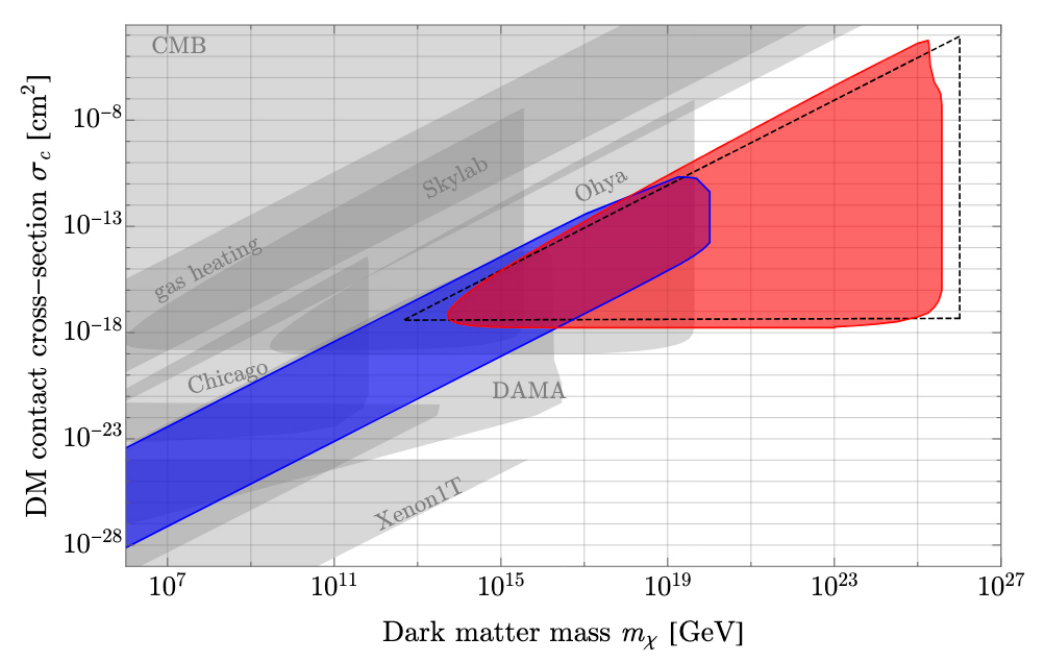}
\end{minipage}
  \caption{\textbf{Etched plastic and ancient mica constraints on very heavy or composite dark matter.}
  Excavated ancient mica limits for the per-nucleon (left) and nucleon-independent (right) parameterizations are shown in blue and red \cite{Acevedo:2021tbl} alongside etched-plastic limits in gray (labeled Skylab and Ohya) from \cite{Bhoonah:2020fys}. Both studies incorporated overburden attenuation and detector thresholds - the features apparent in the red mica bounds result from a detailed Monte Carlo of dark matter attenuation through the Earth's crust \cite{Acevedo:2021tbl}, whereas the plastic etch bounds have more uniform overburden material \cite{Bhoonah:2020fys}. A broader overview of mineral-based searches is given in \cite{Baum:2023cct}.}
  \label{fig:mica-plastic-4panel2}
\end{figure}

Etched plastics and ancient mica now have explored a large swath of parameter space beyond the reach of conventional underground detectors, shown in Figure \ref{fig:mica-plastic-4panel2}. Plastics excel for very large $\sigma_{\rm eff}$ in near-surface conditions, while mica’s immense integration time compensates for the low flux of very heavy candidates. Recent small-scale experimental efforts \cite{Baum:2023cct,Baum:2024eyr} have coalesced into a budding research effort advancing mineral-based astroparticle searches.

\begin{figure}
\centering
\includegraphics[width=0.99\linewidth]{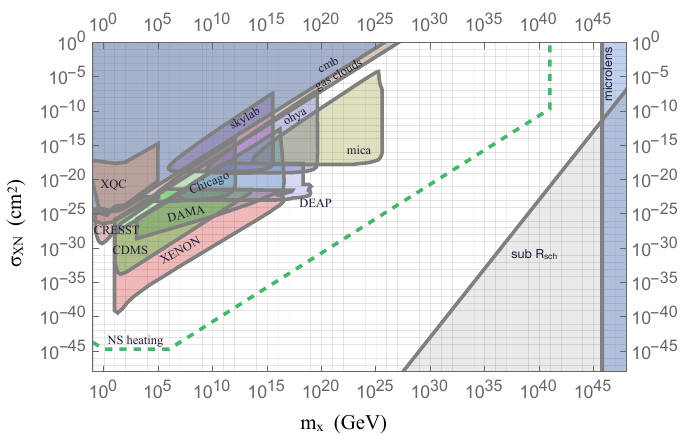}
\caption{
Constraints and sensitivities are shown for dark matter masses 
$m_X = 10^3$--$10^{50}\,\mathrm{GeV}$, for a nucleon-independent geometric cross-section $\sigma_{XN} = \sigma_X$, including representative results 
from underground direct-detection and multiscatter searches, track-etch and 
mica solid-state detectors. The highest masses are constrained by microlensing \cite{Croon:2020wpr}, and the region labelled ``sub $R_{sch}$" indicates objects smaller than their own black hole horizon, which are likely unphysical.
There is also a neutron star dark kinetic-heating prospective sensitivity derived in Ref.~\cite{Baryakhtar:2017dbj}, here adapted for the geometric cross-section, with an upper mass reach requiring that a composite cross a neutron star within 0.1 Gyr. For neutron stars, the energy dark matter deposits falling into the neutron star would be deposited through repeated scatters and observable as a minimum $\sim 1750$ K neutron star temperature, which could be observed as excess infrared emission from a nearby neutron star \cite{Bramante:2024ikc,Raj:2024kjq}.
}
\label{fig:compare_rates_N10}
\end{figure}

It is worth noting that while the current experimental search space, as summarized in Figure \ref{fig:compare_rates_N10}, primarily constrains regions where DM-SM interactions are relatively strong, there is no theoretical prejudice against much weaker interactions. In secluded composite cosmologies, or scenarios involving loosely bound composites and Migdal effect signatures \cite{Acevedo:2021kly,Bleau:2025klr,Acevedo:2024lyr}, the effective coupling to the Standard Model can be extremely tiny, demonstrating that weakly coupled composite dark matter remains a viable possibility.

\section{Astrophysical Searches for Very Heavy Dark Matter}
\label{sec:astrosearches}

Dark matter much heavier than a Planck mass would arrive at Earth with a flux so small that traditional detectors would need to wait centuries for a single encounter. But our cosmos offers vastly larger targets and much longer integration times. Planets, stars, and compact remnants can quietly accumulate dark matter, respond gravitationally to its passage, or have their light lensed in its wake. Such celestial bodies can therefore serve as natural, distributed observatories for very heavy and composite dark matter. 

Heating of planetary interiors, energy release in neutron stars and white dwarfs, and transient effects on background starlight each potentially provide some complementary sensitivity to very heavy dark matter. Moreover, these observables probe distinct aspects of dark matter’s microphysics: stellar capture and thermalization record dark matter's interactions with visible matter; stellar structure constrains its annihilation or decay; and lensing isolates gravitational phenomena, agnostic to coupling details.

Before proceeding to compact star and planetary heating probes, we briefly note some broader constraints and probes of very heavy composite dark matter.
Astrophysical bounds on dark matter self-interactions, reviewed comprehensively in
\cite{Tulin:2017ara}, restrict $\sigma_{\rm self}/m_X \lesssim 1~\mathrm{cm^2/g}$. For the
large composite masses considered throughout this review, this bound is generically satisfied,
since geometric cross-sections scale as $R^2 \propto N^{2/3}$ while composite masses scale
as $N$, so that $\sigma/m$ decreases for larger composites; indeed, self-interaction constraints
were shown to be negligible across the asymmetric fermion-scalar composite parameter space in
\cite{Coskuner:2018are}. Cosmological constraints from thermal history and
$\Delta N_{\rm eff}$ on dark sectors producing light degrees of freedom are model-dependent,
relying on whether ultralight mediators or dark radiation are produced in sufficient abundance
\cite{Cline:2012is,Cline:2013pca,Bleau:2025klr}. Additionally, binary mergers of sufficiently
heavy dark compact objects can produce gravitational wave signals detectable across a range of
frequencies \cite{Diamond:2021dth}, offering a complementary discovery channel for very heavy
composite dark matter.

In the subsections that follow, we review some classes of astrophysical searches.  
First we will examine energy deposition in planetary bodies such as the Earth, where geothermal measurements constrain strongly interacting species.  
Next we briefly review searches using neutron stars and white dwarfs, whose dense bodies act as exquisite calorimeters and accelerators for very heavy dark matter.  
Finally, we will consider gravitational lensing and occultation searches, along with some newly proposed planetary signatures of asteroid size dark matter composites.

\subsection{Earth Heating Searches for Heavy Dark Matter}
\label{subsec:earthheating}

The Earth itself is an enormous calorimeter, with its internal heat budget measured to remarkable precision.  Roughly $40$–$50~\mathrm{TW}$ of heat emerges from the planet’s surface today, sourced primarily by radioactive decay, residual formation energy, and mantle convection.  Any exotic process contributing significantly more than a few terawatts would be in tension with geophysical observations.  This simple accounting has proven a surprisingly powerful probe of dark matter models whose interactions allow efficient capture by the Earth.

As first emphasized in Ref.~\cite{Mack:2007xj}, if dark matter elastically scatters with nuclei, a fraction of the Galactic flux will be gravitationally captured after losing sufficient kinetic energy in transit through the planet.  Subsequent scattering and thermalization bring the captured population to rest in the core, where its density is highest.  In annihilating scenarios, the equilibrium between capture and annihilation determines a steady heating power
\begin{equation}
\dot Q_X \simeq f_{\rm dep}\,m_X\,C_X,
\label{eq:heatflow}
\end{equation}
where $C_X$ is the capture rate and $f_{\rm dep}$ the fraction of annihilation energy deposited locally.  For per-nucleon cross sections $\sigma_{X n}\!\gtrsim\!10^{-38}\,\mathrm{cm^2}$ and masses near the TeV–PeV scale, the heating rate can exceed the measured terrestrial heat flow, excluding such parameters.  This constitutes the canonical ``Earth heat flow bound'' on interacting dark matter.

Subsequent analyses have refined this picture.  
Reference~\cite{Bramante:2019fhi} extended these limits to both annihilating and non-annihilating heavy dark matter, including applications to Mars and other rocky planets.  
They noted that the capture probability saturates for large cross sections—when the mean free path becomes shorter than the planetary radius—and that capture efficiency declines for extremely heavy dark matter whose flux through Earth is low.  
By numerically integrating dark matter trajectories through realistic density and temperature profiles, they derived the differential heating rate in the planetary core, confirming that existing geothermal data exclude large regions of parameter space where captured dark matter would generate tens of terawatts of excess heat.

A qualitatively different constraint arises for non-annihilating or asymmetric dark matter capable of gravitational collapse after accumulation.  
If a captured population of sufficiently massive, self-gravitating dark matter particles forms a black hole, the resulting Hawking radiation could either heat the core or, if unchecked, consume the planet.  
Reference~\cite{Acevedo:2020gro} performed a detailed treatment of this process (see also results for other planets \cite{Croon:2023bmu}), calculating the capture, thermalization, and collapse of asymmetric dark matter in both the Sun and Earth.  
Ref.~\cite{Acevedo:2020gro} showed that for intermediate cross sections and composite dark matter masses up to $\sim10^{39}\,\mathrm{GeV}$, the accumulated core could exceed its Chandrasekhar-like threshold and collapse into an evaporating black hole, injecting substantial thermal energy into the planetary interior, or even cause the Sun or Earth to be entirely swallowed by the black hole formed from non-annihilating dark matter.  
This and the absence of anomalous heat flux from the Earth’s surface therefore sets robust upper limits on the product of capture and dark matter collapse efficiencies.  
In the same work, the authors also derived corresponding constraints from neutrino fluxes produced by Hawking evaporation from smaller black holes formed within the Sun or Earth, further illustrating how planetary and stellar interiors are sensitive to dark matter composites.

Most recently, Ref.~\cite{Cappiello:2025yfe} reexamined the equilibrium assumption underlying earlier heat-flow limits, developing a full thermal evolution model of Earth’s core under dark matter heating.  
By numerically solving the time-dependent heat equation with annihilation energy injection, they showed that dark matter capable of producing the previously allowed $\sim10~\mathrm{TW}$ of heating would in fact melt a large fraction of the solid inner core over geological timescales, conflicting with seismic constraints on the core’s solid structure. This result strengthens Earth-based limits by roughly an order of magnitude and highlight the importance of internal heat transport and equilibration times when applying geophysical constraints to dark matter.

There is also a related search that probed composite dark matter using meteor radar systems.
Reference~\cite{Dhakal:2022rwn} demonstrated that dark matter passing through Earth’s atmosphere could create ionization trails detectable by radar, analogous to meteoroid echoes.

\subsection{Neutron Star, White Dwarf, and Other Signatures of Very Heavy Dark Matter}
\label{subsec:compactstars}

This subsection will be brief, as the subject of dark matter signatures in compact stars has been reviewed recently in~\cite{Bramante:2023djs}.  
Nevertheless, we summarize here the principal mechanisms by which very heavy or composite dark matter could be probed through neutron stars and white dwarfs, emphasizing physical processes that determine their sensitivity.

Compact stars represent the most extreme known baryonic environments, with densities possibly exceeding those of nuclei, relativistic gravitational potentials, and in the case of neutron stars, potentially superfluid cores and intense magnetic fields.  
These conditions render them rather sensitive to dark matter interactions spanning a wide mass range, from sub-GeV to super-Planckian masses, including composite dark matter.  
For very heavy dark matter, the relevant processes can be grouped into four broad mechanisms: capture and associated dark kinetic heating, annihilation and induced reactions, collapse to black holes, and catastrophic stellar disruptions.  
We briefly review each below.

\paragraph{Capture, kinetic, and annihilation heating.}
As dark matter passes through a compact star, scattering on constituent nuclei or leptons can lead to gravitational capture. Captured particles thermalize and settle in the stellar core, converting the majority of their kinetic energy into heat.  
Even for non-annihilating dark matter, this kinetic heating can raise the surface temperature of old neutron stars above the $\sim10^2$~K expected from standard cooling, up to around 1750 K ~\cite{Baryakhtar:2017dbj}, given the flux of dark matter by Earth. Figure \ref{fig:compare_rates_N10} shows the projected sensitivity, in terms of the geometric scattering cross-section, for a nearby $T < 1750$ K neutron star temperature measurement. The large volumetric exposure of a neutron star collecting dark matter over Gyr timescales results in a mass reach shown in Figure \ref{fig:compare_rates_N10} far exceeding terrestrial searches. 

If the dark matter self-annihilates, the captured population will reach an equilibrium between capture and annihilation, producing a steady heating power $\dot{Q}_X = m_X C_X$ analogous to Eq.~(\ref{eq:heatflow}) for Earth heating \cite{Kouvaris:2007ay,Bertone:2007ae}.  
In neutron stars, the resulting thermal energy modifies the neutron star's late-time cooling curve, surface temperatures up to 2500 K~\cite{Baryakhtar:2017dbj}, leading to a number of proposals for expending substantial next-generation telescope time observing nearby pulsars \cite{Bramante:2024ikc,Raj:2024kjq}. Specifically, these proposals for imaging nearby neutron stars include observing campaigns using JWST for higher neutron star temperatures \cite{Raj:2024kjq}, and future thirty meter class telescopes for an ambitious campaign to image known pulsars at a $\sim$ 2000 K sensitivity limit \cite{Bramante:2024ikc}. A related class of models involves inelastic reactions between dark matter and nucleons, including co-annihilation and induced nucleon decay~\cite{McKeen:2021jbh}, each injecting energy or altering chemical composition within the stellar core.

\paragraph{Collapse to black holes.}
For asymmetric or non-annihilating dark matter, continued accumulation can lead to self-gravitational collapse once the enclosed dark core exceeds its Jeans mass.  
The resulting tiny black hole either evaporates via Hawking radiation or grows by accreting surrounding matter, depending on its mass and the density of compact star around it.  
The absence of observed neutron stars in dense dark matter environments thus constrains the rate of such implosions, and in turn imply limits on and detection prospects for heavy asymmetric dark matter ~\cite{Goldman:1989nd,deLavallaz:2010wp,Kouvaris:2010jy,McDermott:2011jp,Bramante:2014zca,Bramante:2016mzo,Bramante:2017ulk,Bhattacharya:2023stq,Basumatary:2024uwo,Bramante:2024idl}.

\paragraph{Thermonuclear and structural effects in white dwarfs.}
In white dwarfs, similar processes may trigger explosive outcomes.  
Dark matter accumulation and self-gravitational collapse can heat a localized region beyond the fusion threshold, igniting runaway nuclear burning and producing a Type~Ia–like supernova~\cite{Acevedo:2020avd,Bramante:2015cua,Fedderke:2019jur,Acevedo:2023cab}.  
Alternatively, if a black hole forms from the captured dark core, it can consume the star on short timescales, converting it into a solar-mass black hole \cite{Acevedo:2019gre,Janish:2019nkk}. In addition, if a large composite dark matter object intercepts a white dwarf, it is possible for it to deposit enough energy to directly cause an explosion during a single passage through the white dwarf \cite{Acevedo:2021kly,Acevedo:2020avd,Fedderke:2019jur,Graham:2018efk,Raj:2023azx}.

\subsection{Lensing and Stellar Occlusion Searches}

Primordial black holes (PBHs) remain the archetypal very heavy dark matter candidate—compact, gravitationally interacting, and long studied as microlensing targets. Their properties and constraints have been extensively reviewed~\cite{Carr:2023tpt}, so we only note here that PBHs first established the paradigm for gravitational detection of compact dark matter through transient amplification of stellar light. That framework now extends to a broader class of dark objects—those that may lens, scatter, or dim starlight as they cross our line of sight.

Reference~\cite{Croon:2020wpr} examined microlensing by extended dark matter structures such as minihalos and boson stars. For lenses with size comparable to or larger than the Einstein radius, the amplification curves deviate from the canonical symmetric brightening of point-like lenses, instead producing shallow or multi-peaked events depending on the internal density profile. Recasting OGLE and EROS data, they showed sensitivity to dark halos over a wide mass range, $10^{-10}$–$10^{2}\,M_\odot$, including configurations that transition smoothly between the compact and diffuse regimes. In addition, dark matter contained in halo-like structures may manifest differently than diffuse dark matter, as explored for neutron star-dark halo interactions in \cite{Edwards:2020afl,Bramante:2021dyx}, which respectively explored axion-miniclusters converting to radio bursts in neutron star magnetic fields, and subhalo dark matter heating of neutron stars.

A complementary signature arises when dark compact objects attenuate rather than amplify light. Reference~\cite{Bramante:2024hbr} proposed that photon-scattering dark matter clumps can transiently \emph{dim} background stars. In this case, the relevant quantity is the (literal) optical depth,
\[
\tau_0 = R_X\, n_X\, \sigma_{X\gamma},
\]
where $\sigma_{X\gamma}$ governs scattering or absorption of photons. For $\tau_0 \!\gtrsim\! 1$, extended dark clouds with radii $R_X \!\sim\! 10$–$10^4\,R_\odot$ can partially occlude stellar light, producing light curves distinguishable from exoplanet transits or dust extinction. Null results from microlensing surveys already constrain such “dark lampshades” over masses $10^{26}$–$10^{35}\,\mathrm{g}$, excluding scenarios where a significant fraction of dark matter resides in photon-interacting composites.

\subsection{Asteroid-like dark matter signatures}
Recent work has shown that planetary surfaces and minor bodies may also serve as archives of very heavy dark matter interactions. 
Ref.~\cite{Picker:2025ofy} showed that dark matter with gram to kiloton mass could leave observable imprints as it collides with asteroids, moons, or planetary rings.  
Such impacts generate localized cratering, melt signatures, or disruptions of ring structure.
Correlating crater morphologies, age distributions, and ring dynamics, \cite{Picker:2025ofy} obtained limits on heavy dark matter fluxes beyond the sensitivity of terrestrial searches. Complementary constraints were proposed in~\cite{DeRocco:2025ovr}, which examined subsurface heating and impact shock signatures on icy moons such as Ganymede.  
In this scenario, the transit of composite dark matter through an ocean-bearing body deposits sufficient energy to induce localized melting, fracturing, and transient vapor plumes possibly detectable by future orbiting satellites.

% Summary Points
\begin{summary}[SUMMARY POINTS]
\begin{enumerate}
\item Multiscatter phenomenology has established a predictive framework for terrestrial detection of dark matter far above the weak scale, with current constraints extending up to $\sim 10^{20}\,\mathrm{GeV}$ from xenon, argon, scintillating, and plastic etch detectors.
\item Composite dark matter models span a continuum from tightly-bound nuclear-scale states to loosely-bound extended structures, with different modes of interaction that are still being explored, including geometric cross sections, interior acceleration, and per-nucleon scattering at large effective cross-sections. These motivate additional research into heavy composite dark matter theory and accompanying searches with both direct detection experiments and astrophysical observations.
\item Geological and astrophysical searches, including mineral searches, neutron star heating, stellar occlusion, and gravitational lensing, for now provide prospective reach for the rarest, heaviest dark matter composites.
\end{enumerate}
\end{summary}

% Future Issues
\begin{issues}[FUTURE ISSUES]
\begin{enumerate}
\item The dynamics and cosmological formation of loosely bound dark composites should be explored in greater detail, linking their structure and cosmological formation to terrestrial detection and astrophysical signatures.
\item Future large-volume detector analyses should incorporate time-correlated multiscatter and track morphologies, to obtain unique sensitivity to very heavy composite dark matter.
\item Mineral and solid-state archives will require additional calibration of track formation and track annealing to more accurately convert observed features into cross-section and flux limits with controlled uncertainties.
\item Additional geological and astrophysical methods for finding the heaviest dark matter candidates should be explored.
\end{enumerate}
\end{issues}

%Disclosure
\section*{DISCLOSURE STATEMENT}
The author is not aware of any affiliations, memberships, funding, or financial holdings that
might be perceived as affecting the objectivity of this review. 

% Acknowledgements
\section*{ACKNOWLEDGMENTS}
The author thanks the anonymous referee for many useful suggestions and clarifications. The author acknowledges support from the Natural Sciences and Engineering Research Council of Canada (NSERC), the Ontario Early Researcher Award (ERA), and the Canada Foundation for Innovation (CFI). This research was undertaken thanks in part to funding from the Arthur B. McDonald Canadian Astroparticle Physics Research Institute.
\bibliographystyle{ar-style5}
\bibliography{refs}

\providecommand{\noopsort}[1]{}\providecommand{\singleletter}[1]{#1}%
\begin{thebibliography}{132}
\expandafter\ifx\csname natexlab\endcsname\relax\def\natexlab#1{#1}\fi

\bibitem{Raffelt:1996wa}
Raffelt GG.
\newblock University of Chicago Press (1996)

\bibitem{Jungman:1995df}
Jungman G, Kamionkowski M, Griest K.
\newblock \textit{Phys.\ Rept.} 267:195--373 (1996)

\bibitem{Witten:1984rs}
Witten E.
\newblock \textit{Phys. Rev. D} 30:272--285 (1984)

\bibitem{Farhi:1984qu}
Farhi E, Jaffe RL.
\newblock \textit{Phys. Rev. D} 30:2379--2390 (1984)

\bibitem{DeRujula:1984axn}
R{\'u}jula AD, Glashow SL.
\newblock \textit{Nature} 312:734--737 (1984)

\bibitem{Ou:2023adg}
Ou X, Eilers AC, Necib L, Frebel A.
\newblock \textit{Mon. Not. Roy. Astron. Soc.} 528(1):693--710 (2024)

\bibitem{Feder:2014dm}
Feder T.
\newblock \textit{Phys. Today} 67(9):9--15 (2014), estimate: LZ \$55 million.

\bibitem{GlobalArgonDarkMatter:2024wtv}
Agnes P, et~al.
\newblock \textit{Front. in Phys.} 12:1387069 (2024)

\bibitem{Akerib:2022ort}
Akerib DS, et~al. 2022.
\newblock In \textit{{Snowmass 2021}}

\bibitem{OECD:2018longview}
{OECD}.
\newblock Paris: Organisation for Economic Co-operation and Development (2018)

\bibitem{Billard:2013qya}
Billard J, Strigari LE, Figueroa-Feliciano E.
\newblock \textit{Phys.\ Rev.\ D} 89:023524 (2014)

\bibitem{OHare:2016rdm}
O’Hare CAJ.
\newblock \textit{Phys.\ Rev.\ D} 94:063527 (2016)

\bibitem{Acevedo:2021tbl}
Acevedo JF, Bramante J, Goodman A.
\newblock \textit{JCAP} 11:085 (2023)

\bibitem{Baum:2023cct}
Baum S, et~al.
\newblock \textit{Phys. Dark Univ.} 41:101245 (2023)

\bibitem{Bramante:2023djs}
Bramante J, Raj N.
\newblock \textit{Phys. Rept.} 1052:1--48 (2024)

\bibitem{Bhoonah:2019eyo}
Bhoonah A, Bramante J, Song N.
\newblock \textit{Phys. Rev. D} 101(5):055040 (2020)

\bibitem{Monteiro:2020wcb}
Monteiro F, Afek G, Carney D, Krnjaic G, Wang J, Moore DC.
\newblock \textit{Phys. Rev. Lett.} 125(18):181102 (2020)

\bibitem{Arvanitaki:2024taq}
Arvanitaki A, Dimopoulos S, Galanis M.
\newblock \textit{Phys. Rev. D} 111(5):055015 (2025)

\bibitem{Bodas:2025vff}
Bodas A, Ghosh S, Harnik R  (2025)

\bibitem{Bramante:2024pyc}
Bramante J, Cappiello CV, Diamond M, Kim JL, Liu Q, Vincent AC.
\newblock \textit{Phys. Rev. D} 110(4):043041 (2024)

\bibitem{Krnjaic:2014xza}
Krnjaic G, Sigurdson K.
\newblock \textit{Phys. Lett.} B751:464--468 (2015)

\bibitem{Detmold:2014qqa}
Detmold W, McCullough M, Pochinsky A.
\newblock \textit{Phys. Rev.} D90(11):115013 (2014)

\bibitem{Detmold:2014kba}
Detmold W, McCullough M, Pochinsky A.
\newblock \textit{Phys. Rev. D} 90(11):114506 (2014)

\bibitem{Wise:2014ola}
Wise MB, Zhang Y.
\newblock \textit{JHEP} 02:023 (2015), [Erratum: JHEP 10, 165 (2015)]

\bibitem{Hardy:2014mqa}
Hardy E, Lasenby R, March-Russell J, West SM.
\newblock \textit{JHEP} 06:011 (2015)

\bibitem{Bramante:2018qbc}
Bramante J, Broerman B, Lang RF, Raj N.
\newblock \textit{Phys. Rev.} D98(8):083516 (2018)

\bibitem{Gresham:2018anj}
Gresham MI, Lou HK, Zurek KM.
\newblock \textit{Phys. Rev.} D98(9):096001 (2018)

\bibitem{Ibe:2018juk}
Ibe M, Kamada A, Kobayashi S, Nakano W.
\newblock \textit{JHEP} 11:203 (2018)

\bibitem{Grabowska:2018lnd}
Grabowska DM, Melia T, Rajendran S.
\newblock \textit{Phys. Rev. D} 98(11):115020 (2018)

\bibitem{Coskuner:2018are}
Coskuner A, Grabowska DM, Knapen S, Zurek KM.
\newblock \textit{Phys. Rev. D} 100(3):035025 (2019)

\bibitem{Bai:2018dxf}
Bai Y, Long AJ, Lu S.
\newblock \textit{Phys. Rev. D} 99(5):055047 (2019)

\bibitem{Bai:2019ogh}
Bai Y, Berger J.
\newblock \textit{JHEP} 05:160 (2020)

\bibitem{Fedderke:2024hfy}
Fedderke MA, Kaplan DE, Mathur A, Rajendran S, Tanin EH.
\newblock \textit{Phys. Rev. D} 109(12):12 (2024)

\bibitem{Wise:2014jva}
Wise MB, Zhang Y.
\newblock \textit{Phys. Rev. D} 90(5):055030 (2014), [Erratum: Phys.Rev.D 91,
  039907 (2015)]

\bibitem{Gresham:2017cvl}
Gresham MI, Lou HK, Zurek KM.
\newblock \textit{Phys. Rev. D} 97(3):036003 (2018)

\bibitem{Gresham:2017zqi}
Gresham MI, Lou HK, Zurek KM.
\newblock \textit{Phys. Rev.} D96(9):096012 (2017)

\bibitem{Acevedo:2021kly}
Acevedo JF, Bramante J, Goodman A.
\newblock \textit{Phys. Rev. D} 105(2):023012 (2022)

\bibitem{Acevedo:2024lyr}
Acevedo JF, Boukhtouchen Y, Bramante J, Cappiello C, Mohlabeng G, Tyagi N.
\newblock \textit{JCAP} 03:013 (2025)

\bibitem{Bleau:2025klr}
Bleau K, Boukhtouchen Y, Bramante J, Kulkarni R  (2025)

\bibitem{Liddle:1992fmk}
Liddle AR, Madsen MS.
\newblock \textit{Int. J. Mod. Phys. D} 1:101--144 (1992)

\bibitem{Liebling:2012fv}
Liebling SL, Palenzuela C.
\newblock \textit{Living Rev. Rel.} 26(1):1 (2023)

\bibitem{Giudice:2016zpa}
Giudice GF, McCullough M, Urbano A.
\newblock \textit{JCAP} 10:001 (2016)

\bibitem{Hippert:2022snq}
Hippert M, Dillingham E, Tan H, Curtin D, Noronha-Hostler J, Yunes N.
\newblock \textit{Phys. Rev. D} 107(11):115028 (2023)

\bibitem{Fan:2013yva}
Fan J, Katz A, Randall L, Reece M.
\newblock \textit{Phys. Dark Univ.} 2:139--156 (2013)

\bibitem{Foot:2013vna}
Foot R.
\newblock \textit{Phys. Rev. D} 88(2):023520 (2013)

\bibitem{Erickcek:2011us}
Erickcek AL, Sigurdson K.
\newblock \textit{Phys. Rev. D} 84:083503 (2011)

\bibitem{Buckley:2017ttd}
Buckley MR, DiFranzo A.
\newblock \textit{Phys. Rev. Lett.} 120(5):051102 (2018)

\bibitem{Chang:2018bgx}
Chang S, Escudero M, Egana-Ugrinovic JFG, Essig R, Kouvaris C.
\newblock \textit{Phys. Rev. D} 99(6):063005 (2019)

\bibitem{Ryan:2021dis}
Ryan M, Gurian J, Shandera S, Jeong D.
\newblock \textit{Astrophys. J.} 934:120 (2022)

\bibitem{Gurian:2022nbx}
Gurian J, Ryan M, Schon S, Jeong D, Shandera S.
\newblock \textit{Astrophys. J. Lett.} 939(1):L12 (2022), [Erratum:
  Astrophys.J.Lett. 949, L44 (2023), Erratum: Astrophys.J. 949, L44 (2023)]

\bibitem{Bramante:2023ddr}
Bramante J, Diamond M, Kim JL.
\newblock \textit{JCAP} 02:002 (2024)

\bibitem{Gemmell:2023trd}
Gemmell C, Roy S, Shen X, Curtin D, Lisanti M, et~al.
\newblock \textit{Astrophys. J.} 967(1):21 (2024)

\bibitem{Roy:2024bcu}
Roy S, Shen X, Barron J, Lisanti M, Curtin D, et~al.
\newblock \textit{Astrophys. J.} 982(2):175 (2025)

\bibitem{Lu:2024xnb}
Lu Y, Picker ZSC, Profumo S, Kusenko A.
\newblock \textit{Phys. Rev. D} 111(4):043005 (2025)

\bibitem{Baker:2019ndr}
Baker MJ, Kopp J, Long AJ.
\newblock \textit{Phys. Rev. Lett.} 125(15):151102 (2020)

\bibitem{Baker:2021nyl}
Baker MJ, Breitbach M, Kopp J, Mittnacht L.
\newblock \textit{Phys. Lett. B} 868:139625 (2025)

\bibitem{Asadi:2021pwo}
Asadi P, Kramer ED, Kuflik E, Ridgway GW, Slatyer TR, Smirnov J.
\newblock \textit{Phys.\ Rev.\ D} 104(9):095013 (2021), iNSPIRE-HEP:1852316

\bibitem{Asadi:2022vkc}
Asadi P, Kramer ED, Kuflik E, Slatyer TR, Smirnov J.
\newblock \textit{JHEP} 07:006 (2022)

\bibitem{Bai:2018dqn}
Bai Y, Long AJ, Lu S.
\newblock \textit{Phys.\ Rev.\ D} 99(5):055047 (2019), iNSPIRE-HEP:1697698

\bibitem{Gross:2021qgx}
Gross C, Landini G, Strumia A, Teresi D.
\newblock \textit{J. High Energ. Phys.} 2021(9):33 (2021), iNSPIRE-HEP:1862346

\bibitem{Kusenko:1997si}
Kusenko A, Shaposhnikov ME.
\newblock \textit{Phys. Lett. B} 418:46--54 (1998)

\bibitem{Amin:2019ums}
Amin MA, Mocz P.
\newblock \textit{Phys. Rev. D} 100(6):063507 (2019)

\bibitem{Hong:2020est}
Hong JP, Jung S, Xie KP.
\newblock \textit{Phys. Rev. D} 102(7):075028 (2020)

\bibitem{An:2022toi}
An H, Tong X, Zhou S.
\newblock \textit{Phys. Rev. D} 107(2):023522 (2023)

\bibitem{Freese:2023fcr}
Freese K, Winkler MW.
\newblock \textit{Phys. Rev. D} 107(8):083522 (2023)

\bibitem{Baldes:2018emh}
Baldes I, Garcia-Cely C.
\newblock \textit{JHEP} 05:190 (2019)

\bibitem{Schwaller:2015tja}
Schwaller P.
\newblock \textit{Phys. Rev. Lett.} 115(18):181101 (2015)

\bibitem{Davoudiasl:2019xeb}
Davoudiasl H, Mohlabeng G.
\newblock \textit{JHEP} 04:177 (2020)

\bibitem{Bhoonah:2020oov}
Bhoonah A, Bramante J, Nerval S, Song N.
\newblock \textit{JCAP} 04:043 (2021)

\bibitem{Bernabei:1999ui}
Bernabei R, et~al.
\newblock \textit{Phys. Rev. Lett.} 83:4918--4921 (1999)

\bibitem{Bramante:2018tos}
Bramante J, Broerman B, Kumar J, Lang RF, Pospelov M, Raj N.
\newblock \textit{Phys. Rev. D} 99(8):083010 (2019)

\bibitem{Cappiello:2020lbk}
Cappiello CV, Collar JI, Beacom JF.
\newblock \textit{Phys. Rev. D} 103(2):023019 (2021)

\bibitem{Digman:2019wdm}
Digman MC, Cappiello CV, Beacom JF, Hirata CM, Peter AHG.
\newblock \textit{Phys. Rev. D} 100(6):063013 (2019), [Erratum: Phys.Rev.D 106,
  089902 (2022)]

\bibitem{Jacobs:2014yca}
Jacobs DM, Starkman GD, Lynn BW.
\newblock \textit{Mon. Not. Roy. Astron. Soc.} 450(4):3418--3430 (2015)

\bibitem{Bramante:2019yss}
Bramante J, Kumar J, Raj N.
\newblock \textit{Phys. Rev. D} 100(12):123016 (2019)

\bibitem{DEAPCollaboration:2021raj}
Adhikari P, et~al.
\newblock \textit{Phys. Rev. Lett.} 128(1):011801 (2022)

\bibitem{Acevedo:2020avd}
Acevedo JF, Bramante J, Goodman A.
\newblock \textit{Phys. Rev. D} 103(12):123022 (2021)

\bibitem{Ebadi:2021cte}
Ebadi R, et~al.
\newblock \textit{Phys. Rev. D} 104(1):015041 (2021)

\bibitem{Drukier:2018pdy}
Drukier AK, Baum S, Freese K, G\'orski M, Stengel P.
\newblock \textit{Phys. Rev. D} 99(4):043014 (2019)

\bibitem{Fung:2025cub}
Fung A, Lucas T, Balogh L, Leybourne M, Vincent AC.
\newblock \textit{Phys. Rev. D} 112(4):043040 (2025)

\bibitem{XENON:2023iku}
Aprile E, et~al.
\newblock \textit{Phys. Rev. Lett.} 130(26):261002 (2023)

\bibitem{LZ:2024psa}
Aalbers J, et~al.
\newblock \textit{Phys. Rev. D} 109(11):112010 (2024)

\bibitem{PandaX:2024qfu}
Bo Z, et~al.
\newblock \textit{Phys. Rev. Lett.} 134(1):011805 (2025)

\bibitem{Bhoonah:2020dzs}
Bhoonah A, Bramante J, Schon S, Song N.
\newblock \textit{Phys. Rev. D} 103(12):123026 (2021)

\bibitem{IceCube:2014xnp}
Aartsen MG, et~al.
\newblock \textit{Eur. Phys. J. C} 74(7):2938 (2014), [Erratum: Eur.Phys.J.C
  79, 124 (2019)]

\bibitem{Bhoonah:2020fys}
Bhoonah A, Bramante J, Courtman B, Song N.
\newblock \textit{Phys. Rev. D} 103(10):103001 (2021)

\bibitem{Skylab}
{Shirk} EK, {Price} PB.
\newblock \textit{The Astrophysical Journal} 220:719--733 (1978)

\bibitem{Orito:1990ny}
Orito S, et~al.
\newblock \textit{Phys. Rev. Lett.} 66:1951--1954 (1991)

\bibitem{Starkman:1990nj}
Starkman GD, Gould A, Esmailzadeh R, Dimopoulos S.
\newblock \textit{Phys. Rev. D} 41:3594 (1990)

\bibitem{Price:1986ky}
Price P, Salamon M.
\newblock \textit{Phys. Rev. Lett.} 56:1226--1229 (1986)

\bibitem{SnowdenIfft:1995ke}
Snowden-Ifft D, Freeman E, Price P.
\newblock \textit{Phys. Rev. Lett.} 74:4133--4136 (1995)

\bibitem{Baum:2024eyr}
Baum S, et~al. 2024

\bibitem{Croon:2020wpr}
Croon D, McKeen D, Raj N.
\newblock \textit{Phys. Rev. D} 101(8):083013 (2020)

\bibitem{Baryakhtar:2017dbj}
Baryakhtar M, Bramante J, Li SW, Linden T, Raj N.
\newblock \textit{Phys. Rev. Lett.} 119(13):131801 (2017)

\bibitem{Bramante:2024ikc}
Bramante J, Mack K, Raj N, Shao L, Tyagi N  (2024)

\bibitem{Raj:2024kjq}
Raj N, Shivanna P, Rachh GN.
\newblock \textit{Phys. Rev. D} 109(12):123040 (2024)

\bibitem{Tulin:2017ara}
Tulin S, Yu HB.
\newblock \textit{Phys. Rept.} 730:1--57 (2018)

\bibitem{Cline:2012is}
Cline JM, Liu Z, Xue W.
\newblock \textit{Phys. Rev. D} 85:101302 (2012)

\bibitem{Cline:2013pca}
Cline JM, Liu Z, Moore G, Xue W.
\newblock \textit{Phys. Rev. D} 89(4):043514 (2014)

\bibitem{Diamond:2021dth}
Diamond MD, Kaplan DE, Rajendran S.
\newblock \textit{JHEP} 01:136 (2023)

\bibitem{Mack:2007xj}
Mack GD, Beacom JF, Bertone G.
\newblock \textit{Phys. Rev. D} 76:043523 (2007)

\bibitem{Bramante:2019fhi}
Bramante J, Buchanan A, Goodman A, Lodhi E.
\newblock \textit{Phys. Rev.} D101(4):043001 (2020)

\bibitem{Acevedo:2020gro}
Acevedo JF, Bramante J, Goodman A, Kopp J, Opferkuch T.
\newblock \textit{JCAP} 04:026 (2021)

\bibitem{Croon:2023bmu}
Croon D, Smirnov J.
\newblock \textit{JCAP} 11:046 (2024)

\bibitem{Cappiello:2025yfe}
Cappiello C, Daylan T.
\newblock \textit{Phys. Rev. D} 112(7):075018 (2025)

\bibitem{Dhakal:2022rwn}
Dhakal P, Prohira S, Cappiello CV, Beacom JF, Palo S, Marino J.
\newblock \textit{Phys. Rev. D} 107(4):043026 (2023)

\bibitem{Kouvaris:2007ay}
Kouvaris C.
\newblock \textit{Phys. Rev. D} 77:023006 (2008)

\bibitem{Bertone:2007ae}
Bertone G, Fairbairn M.
\newblock \textit{Phys. Rev. D} 77:043515 (2008)

\bibitem{McKeen:2021jbh}
McKeen D, Pospelov M, Raj N.
\newblock \textit{Phys. Rev. Lett.} 127(6):061805 (2021)

\bibitem{Goldman:1989nd}
Goldman I, Nussinov S.
\newblock \textit{Phys. Rev.} D40:3221--3230 (1989)

\bibitem{deLavallaz:2010wp}
de~Lavallaz A, Fairbairn M.
\newblock \textit{Phys. Rev.} D81:123521 (2010)

\bibitem{Kouvaris:2010jy}
Kouvaris C, Tinyakov P.
\newblock \textit{Phys. Rev.} D83:083512 (2011)

\bibitem{McDermott:2011jp}
McDermott SD, Yu HB, Zurek KM.
\newblock \textit{Phys. Rev. D} 85:023519 (2012)

\bibitem{Bramante:2014zca}
Bramante J, Linden T.
\newblock \textit{Phys. Rev. Lett.} 113(19):191301 (2014)

\bibitem{Bramante:2016mzo}
Bramante J, Linden T.
\newblock \textit{Astrophys. J.} 826(1):57 (2016)

\bibitem{Bramante:2017ulk}
Bramante J, Linden T, Tsai YD.
\newblock \textit{Phys. Rev. D} 97(5):055016 (2018)

\bibitem{Bhattacharya:2023stq}
Bhattacharya S, Dasgupta B, Laha R, Ray A.
\newblock \textit{Phys. Rev. Lett.} 131(9):091401 (2023)

\bibitem{Basumatary:2024uwo}
Basumatary U, Raj N, Ray A.
\newblock \textit{Phys. Rev. D} 111(4):L041306 (2025)

\bibitem{Bramante:2024idl}
Bramante J, Raj N.
\newblock \textit{Phys. Rev. D} 110(4):043537 (2024)

\bibitem{Bramante:2015cua}
Bramante J.
\newblock \textit{Phys. Rev. Lett.} 115(14):141301 (2015)

\bibitem{Fedderke:2019jur}
Fedderke MA, Graham PW, Rajendran S.
\newblock \textit{Phys. Rev. D} 101(11):115021 (2020)

\bibitem{Acevedo:2023cab}
Acevedo JF, An H, Boukhtouchen Y, Bramante J, Richardson MLA, Sansom L.
\newblock \textit{Phys. Rev. D} 110(8):083004 (2024)

\bibitem{Acevedo:2019gre}
Acevedo JF, Bramante J.
\newblock \textit{Phys. Rev.} D100(4):043020 (2019)

\bibitem{Janish:2019nkk}
Janish R, Narayan V, Riggins P.
\newblock \textit{Phys. Rev. D} 100(3):035008 (2019)

\bibitem{Graham:2018efk}
Graham PW, Janish R, Narayan V, Rajendran S, Riggins P.
\newblock \textit{Phys. Rev. D} 98(11):115027 (2018)

\bibitem{Raj:2023azx}
Raj N.
\newblock \textit{Phys. Rev. D} 109(12):123020 (2024)

\bibitem{Carr:2023tpt}
Carr B, Clesse S, Garcia-Bellido J, Hawkins M, Kuhnel F.
\newblock \textit{Phys. Rept.} 1054:1--68 (2024)

\bibitem{Edwards:2020afl}
Edwards TDP, Kavanagh BJ, Visinelli L, Weniger C.
\newblock \textit{Phys. Rev. Lett.} 127(13):131103 (2021)

\bibitem{Bramante:2021dyx}
Bramante J, Kavanagh BJ, Raj N.
\newblock \textit{Phys. Rev. Lett.} 128(23):231801 (2022)

\bibitem{Bramante:2024hbr}
Bramante J, Diamond MD, Kim JL.
\newblock \textit{Phys. Rev. Lett.} 134(14):141001 (2025)

\bibitem{Picker:2025ofy}
Picker ZSC.
\newblock \textit{Phys. Rev. D} 112(4):043028 (2025)

\bibitem{DeRocco:2025ovr}
DeRocco W.
\newblock \textit{Phys. Rev. D} 112(11):115023 (2025)

\end{thebibliography}

\end{document}